\documentclass[10pt,twocolumn,twoside]{IEEEtran} 
\usepackage{amssymb}
\usepackage{graphicx,subfigure}
\usepackage[centertags]{amsmath}
\usepackage{epsfig,algorithm,amsmath,amssymb,color,subfigure,url}
\newtheorem{theorem}{Theorem}
\newtheorem{lemma}{Lemma}
\newtheorem{claim}{Claim}
\newtheorem{corollary}{Corollary}
\newtheorem{definition}{Definition}
\long\def\comment#1{}

\begin{document}
\renewcommand{\textfraction}{0}

\title{LS-CS-residual (LS-CS): Compressive Sensing on Least Squares Residual}

\author{Namrata Vaswani
\thanks{N. Vaswani is with the ECE dept at Iowa State University, Ames, IA (email: namrata@iastate.edu).
This research was partially supported by NSF grants ECCS-0725849 and CCF-0917015. A part of this work appeared in \cite{kfcsicip, kfcspap}.
}
}
\date{}
\maketitle \pagestyle{plain} 


\setlength{\arraycolsep}{0.03cm}
\newcommand{\xhat}{\hat{x}}
\newcommand{\xpred}{\hat{x}_{t|t-1}}
\newcommand{\Ppred}{P_{t|t-1}}
\newcommand{\ty}{\tilde{y}_t}
\newcommand{\tty}{\tilde{y}_{t,\text{res}}}
\newcommand{\tw}{\tilde{w}_t}
\newcommand{\ttw}{\tilde{w}_{t,f}}
\newcommand{\betahat}{\hat{\beta}}

\newcommand{\ypast}{y_{1:t-1}}
\newcommand{\sone}{S_{*}}
\newcommand{\sinf}{{S_{**}}}
\newcommand{\smax}{S_{\max}}
\newcommand{\smin}{S_{\min}}
\newcommand{\samax}{S_{a,\max}}
\newcommand{\Nhat}{{\hat{N}}}

\newcommand{\Dnum}{D_{num}}
\newcommand{\pss}{p^{**,i}}
\newcommand{\fr}{f_{r}^i}

\newcommand{\A}{{\cal A}}
\newcommand{\Z}{{\cal Z}}
\newcommand{\B}{{\cal B}}
\newcommand{\R}{{\cal R}}
\newcommand{\reg}{{\cal G}}
\newcommand{\const}{\mbox{const}}

\newcommand{\trace}{\mbox{tr}}

\newcommand{\hsim}{{\hspace{0.0cm} \sim  \hspace{0.0cm}}}
\newcommand{\he}{{\hspace{0.0cm} =  \hspace{0.0cm}}}

\newcommand{\vect}[2]{\left[\begin{array}{cccccc}
     #1 \\
     #2
   \end{array}
  \right]
  }

\newcommand{\matr}[2]{ \left[\begin{array}{cc}
     #1 \\
     #2
   \end{array}
  \right]
  }
\newcommand{\vc}[2]{\left[\begin{array}{c}
     #1 \\
     #2
   \end{array}
  \right]
  }

\newcommand{\gdot}{\dot{g}}
\newcommand{\Cdot}{\dot{C}}
\newcommand{\re}{\mathbb{R}}
\newcommand{\n}{{\cal N}}  
\newcommand{\N}{{\overrightarrow{\bf N}}}  
\newcommand{\chat}{\tilde{C}_t}
\newcommand{\chati}{\chat^i}

\newcommand{\cmin}{C^*_{min}}
\newcommand{\twi}{\tilde{w}_t^{(i)}}
\newcommand{\twj}{\tilde{w}_t^{(j)}}
\newcommand{\wi}{{w}_t^{(i)}}
\newcommand{\twio}{\tilde{w}_{t-1}^{(i)}}

\newcommand{\tWi}{\tilde{W}_n^{(m)}}
\newcommand{\tWj}{\tilde{W}_n^{(k)}}
\newcommand{\Wi}{{W}_n^{(m)}}
\newcommand{\tWio}{\tilde{W}_{n-1}^{(m)}}

\newcommand{\ds}{\displaystyle}

\newcommand{\SAR}{S$\!$A$\!$R }
\newcommand{\MAR}{MAR}
\newcommand{\MMRF}{MMRF}
\newcommand{\AR}{A$\!$R }
\newcommand{\GMRF}{G$\!$M$\!$R$\!$F }
\newcommand{\DTM}{D$\!$T$\!$M }
\newcommand{\MSE}{M$\!$S$\!$E }
\newcommand{\RCS}{R$\!$C$\!$S }
\newcommand{\uomega}{\underline{\omega}}
\newcommand{\y}{v}
\newcommand{\x}{w}
\newcommand{\lu}{\mu}
\newcommand{\g}{g}
\newcommand{\s}{{\bf s}}
\newcommand{\bft}{{\bf t}}
\newcommand{\refmap}{{\cal R}}
\newcommand{\totrefl}{{\cal E}}
\newcommand{\beq}{\begin{equation}}
\newcommand{\eeq}{\end{equation}}
\newcommand{\bdm}{\begin{displaymath}}
\newcommand{\edm}{\end{displaymath}}
\newcommand{\hatz}{\hat{z}}
\newcommand{\hatu}{\hat{u}}
\newcommand{\tilz}{\tilde{z}}
\newcommand{\tilu}{\tilde{u}}
\newcommand{\hhatz}{\hat{\hat{z}}}
\newcommand{\hhatu}{\hat{\hat{u}}}
\newcommand{\tilc}{\tilde{C}}
\newcommand{\hatc}{\hat{C}}
\newcommand{\tim}{n}

\newcommand{\ssp}{\renewcommand{\baselinestretch}{1.0}}
\newcommand{\defd}{\mbox{$\stackrel{\mbox{$\triangle$}}{=}$}}
\newcommand{\goes}{\rightarrow}
\newcommand{\tends}{\rightarrow}
\newcommand{\defn}{\triangleq} 
\newcommand{\se}{&=&}
\newcommand{\sdefn}{& \defn  &}
\newcommand{\sle}{& \le &}
\newcommand{\sge}{& \ge &}
\newcommand{\plusminus}{\stackrel{+}{-}}
\newcommand{\Ey}{E_{Y_{1:t}}}
\newcommand{\ey}{E_{Y_{1:t}}}

\newcommand{\equivto}{\mbox{~~~which is equivalent to~~~}}
\newcommand{\nonzero}{i:\pi^n(x^{(i)})>0}
\newcommand{\nonzeroc}{i:c(x^{(i)})>0}

\newcommand{\supn}{\sup_{\phi:||\phi||_\infty \le 1}}
\newtheorem{remark}{Remark}
\newtheorem{example}{Example}
\newtheorem{ass}{Assumption}
\newtheorem{proposition}{Proposition}

\newtheorem{fact}{Fact}
\newtheorem{heuristic}{Heuristic}
\newcommand{\eps}{\epsilon}
\newcommand{\bd}{\begin{definition}}
\newcommand{\ed}{\end{definition}}
\newcommand{\udq}{\underline{D_Q}}
\newcommand{\td}{\tilde{D}}
\newcommand{\epsinv}{\epsilon_{inv}}
\newcommand{\al}{\mathcal{A}}

\newcommand{\bfx} {\bf X}
\newcommand{\bfy} {\bf Y}
\newcommand{\bfz} {\bf Z}
\newcommand{\ddas}{\mbox{${d_1}^2({\bf X})$}}
\newcommand{\ddbs}{\mbox{${d_2}^2({\bfx})$}}
\newcommand{\dda}{\mbox{$d_1(\bfx)$}}
\newcommand{\ddb}{\mbox{$d_2(\bfx)$}}
\newcommand{\xinc}{{\bfx} \in \mbox{$C_1$}}
\newcommand{\eqa}{\stackrel{(a)}{=}}
\newcommand{\eqb}{\stackrel{(b)}{=}}
\newcommand{\eqe}{\stackrel{(e)}{=}}
\newcommand{\leqc}{\stackrel{(c)}{\le}}
\newcommand{\leqd}{\stackrel{(d)}{\le}}

\newcommand{\leqa}{\stackrel{(a)}{\le}}
\newcommand{\leqb}{\stackrel{(b)}{\le}}
\newcommand{\leqe}{\stackrel{(e)}{\le}}
\newcommand{\leqf}{\stackrel{(f)}{\le}}
\newcommand{\leqg}{\stackrel{(g)}{\le}}
\newcommand{\leqh}{\stackrel{(h)}{\le}}
\newcommand{\leqi}{\stackrel{(i)}{\le}}
\newcommand{\leqj}{\stackrel{(j)}{\le}}

\newcommand{\w}{{W^{LDA}}}
\newcommand{\halpha}{\hat{\alpha}}
\newcommand{\hsigma}{\hat{\sigma}}
\newcommand{\slmax}{\sqrt{\lambda_{max}}}
\newcommand{\slmin}{\sqrt{\lambda_{min}}}
\newcommand{\lmax}{\lambda_{max}}
\newcommand{\lmin}{\lambda_{min}}

\newcommand{\da} {\frac{\alpha}{\sigma}}
\newcommand{\chka} {\frac{\check{\alpha}}{\check{\sigma}}}
\newcommand{\sumo}{\sum _{\underline{\omega} \in \Omega}}
\newcommand{\distance}{d\{(\hatz _x, \hatz _y),(\tilz _x, \tilz _y)\}}
\newcommand{\col}{{\rm col}}
\newcommand{\rcs}{\sigma_0}
\newcommand{\CalR}{{\cal R}}
\newcommand{\df}{{\delta p}}
\newcommand{\dq}{{\delta q}}
\newcommand{\dZ}{{\delta Z}}
\newcommand{\pprime}{{\prime\prime}}

\newcommand{\vn}{N}

\newcommand{\bv}{\begin{vugraph}}
\newcommand{\ev}{\end{vugraph}}
\newcommand{\bi}{\begin{itemize}}
\newcommand{\ei}{\end{itemize}}
\newcommand{\ben}{\begin{enumerate}}
\newcommand{\een}{\end{enumerate}}
\newcommand{\be}{\protect\[}
\newcommand{\ee}{\protect\]}
\newcommand{\bean}{\begin{eqnarray*} }
\newcommand{\eean}{\end{eqnarray*} }
\newcommand{\bea}{\begin{eqnarray} }
\newcommand{\eea}{\end{eqnarray} }
\newcommand{\nn}{\nonumber}
\newcommand{\ba}{\begin{array} }
\newcommand{\ea}{\end{array} }
\newcommand{\ep}{\mbox{\boldmath $\epsilon$}}
\newcommand{\epp}{\mbox{\boldmath $\epsilon '$}}
\newcommand{\Lep}{\mbox{\LARGE $\epsilon_2$}}
\newcommand{\und}{\underline}
\newcommand{\pdif}[2]{\frac{\partial #1}{\partial #2}}
\newcommand{\odif}[2]{\frac{d #1}{d #2}}
\newcommand{\dt}[1]{\pdif{#1}{t}}
\newcommand{\urho}{\underline{\rho}}

\newcommand{\spc}{{\cal S}}
\newcommand{\tspc}{{\cal TS}}

\newcommand{\uv}{\underline{v}}
\newcommand{\us}{\underline{s}}
\newcommand{\uc}{\underline{c}}
\newcommand{\utheta}{\underline{\theta}^*}
\newcommand{\ualpha}{\underline{\alpha^*}}

\newcommand{\uxy}{\underline{x}^*}
\newcommand{\uxyj}{[x^{*}_j,y^{*}_j]}
\newcommand{\arcl}[1]{arclen(#1)}
\newcommand{\one}{{\mathbf{1}}}

\newcommand{\uxyjt}{\uxy_{j,t}}
\newcommand{\E}{\mathbb{E}}

\newcommand{\rhomat}{\left[\begin{array}{c}
                        \rho_3 \ \rho_4 \\
                        \rho_5 \ \rho_6
                        \end{array}
                   \right]}
\newcommand{\deltat}{\tau} 
\newcommand{\deltatt}{\Delta t_1}
\newcommand{\ceil}[1]{\ulcorner #1 \urcorner}

\newcommand{\xxi}{x^{(i)}}
\newcommand{\txi}{\tilde{x}^{(i)}}
\newcommand{\txj}{\tilde{x}^{(j)}}

\newcommand{\mi}[1]{{#1}^{(m,i)}}

\setlength{\arraycolsep}{0.05cm}

\newcommand{\Section}[1]{ \section{#1} } 
\newcommand{\Subsection}[1]{  \subsection{#1}   } 
\newcommand{\Subsubsection}[1]{   \subsubsection{#1} } 

\newcommand{\rest}{{T_\text{rest}}}
\newcommand{\zetahat}{\hat{\zeta}}
\newcommand{\sm}{\text{small}}
\newcommand{\tDelta}{{\tilde{\Delta}}}
\newcommand{\tDeltae}{{\tilde{\Delta}_e}}
\newcommand{\tT}{{\tilde{T}}}
\newcommand{\add}{{\cal A}}
\newcommand{\rem}{{\cal R}}
\newtheorem{sigmodel}{Signal Model}

\newcommand{\thr}{{\text{thr}}}
\newcommand{\delthr}{{\text{del-thr}}}
\newcommand{\delbound}{{b}}
\newcommand{\err}{{\text{err}}}
\newcommand{\Q}{{\cal Q}}

\newcommand{\dett}{{\text{det}}}
\newcommand{\CSres}{{\text{CSres}}}
\newcommand{\diff}{{\text{diff}}}


\begin{abstract}
We consider the problem of recursively and causally reconstructing time sequences of sparse signals (with unknown and time-varying sparsity patterns) from a limited number of noisy linear measurements. The sparsity pattern is assumed to change slowly with time. The idea of our proposed solution, LS-CS-residual (LS-CS), is to replace compressed sensing (CS) on the observation by CS on the least squares (LS) residual computed using the previous estimate of the support. We bound CS-residual error and show that when the number of available measurements is small, the bound is much smaller than that on CS error if the sparsity pattern changes slowly enough. We also obtain conditions for ``stability" of LS-CS over time for a signal model that allows support additions and removals, and that allows coefficients to gradually increase (decrease) until they reach a constant value (become zero). By ``stability", we mean that the number of misses and extras in the support estimate remain bounded by time-invariant values (in turn implying a time-invariant bound on LS-CS error). The concept is meaningful only if the bounds are small compared to the support size. Numerical experiments backing our claims are shown (also include a dynamic MRI example).%
\end{abstract}



\begin{figure}
\centerline{
\subfigure[Top: larynx image sequence, Bottom: cardiac sequence]{
\label{examples}
\begin{tabular}{c}
\epsfig{file = 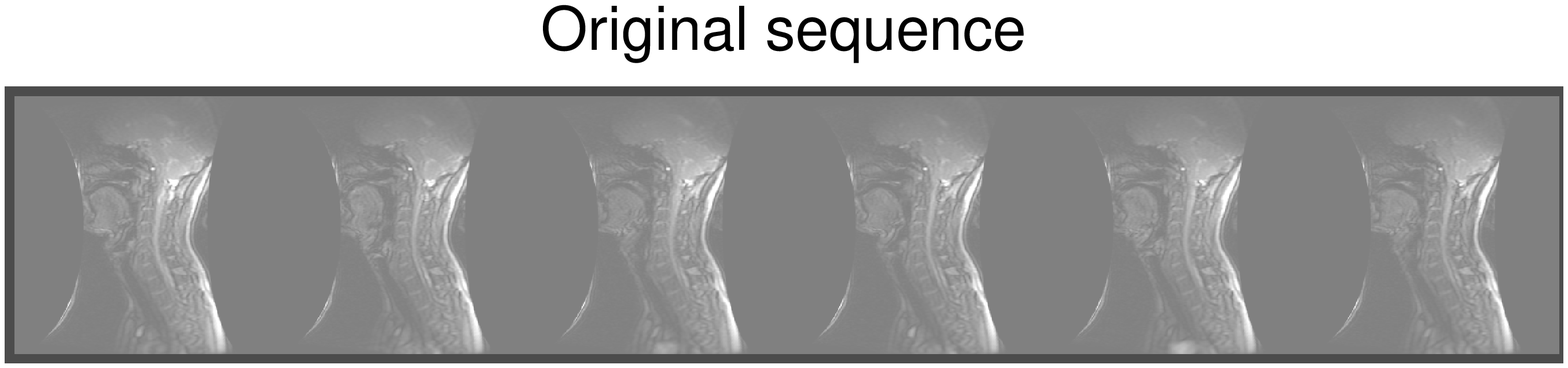,height=2cm} \vspace{-0.2in}  \\ 
\epsfig{file = 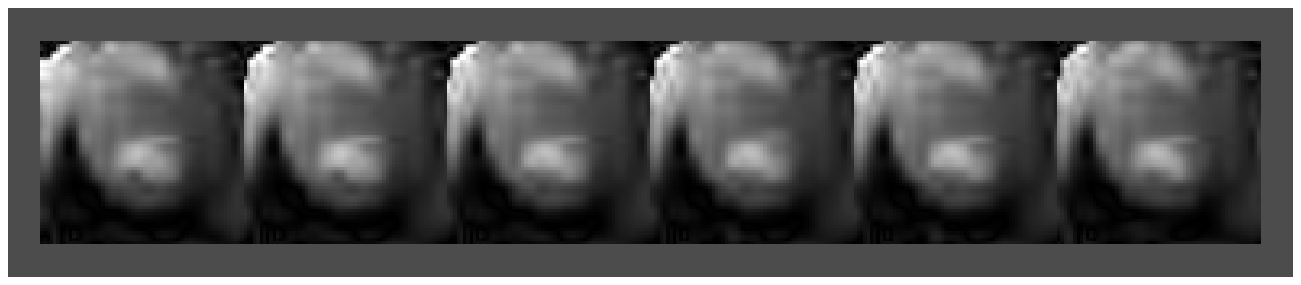, height=1.5cm} 
\end{tabular}
}
}
\centerline{
\subfigure[Slow support change plots. Left: additions, Right: removals]{
\label{supp90_99}
\begin{tabular}{cc}
\epsfig{file = 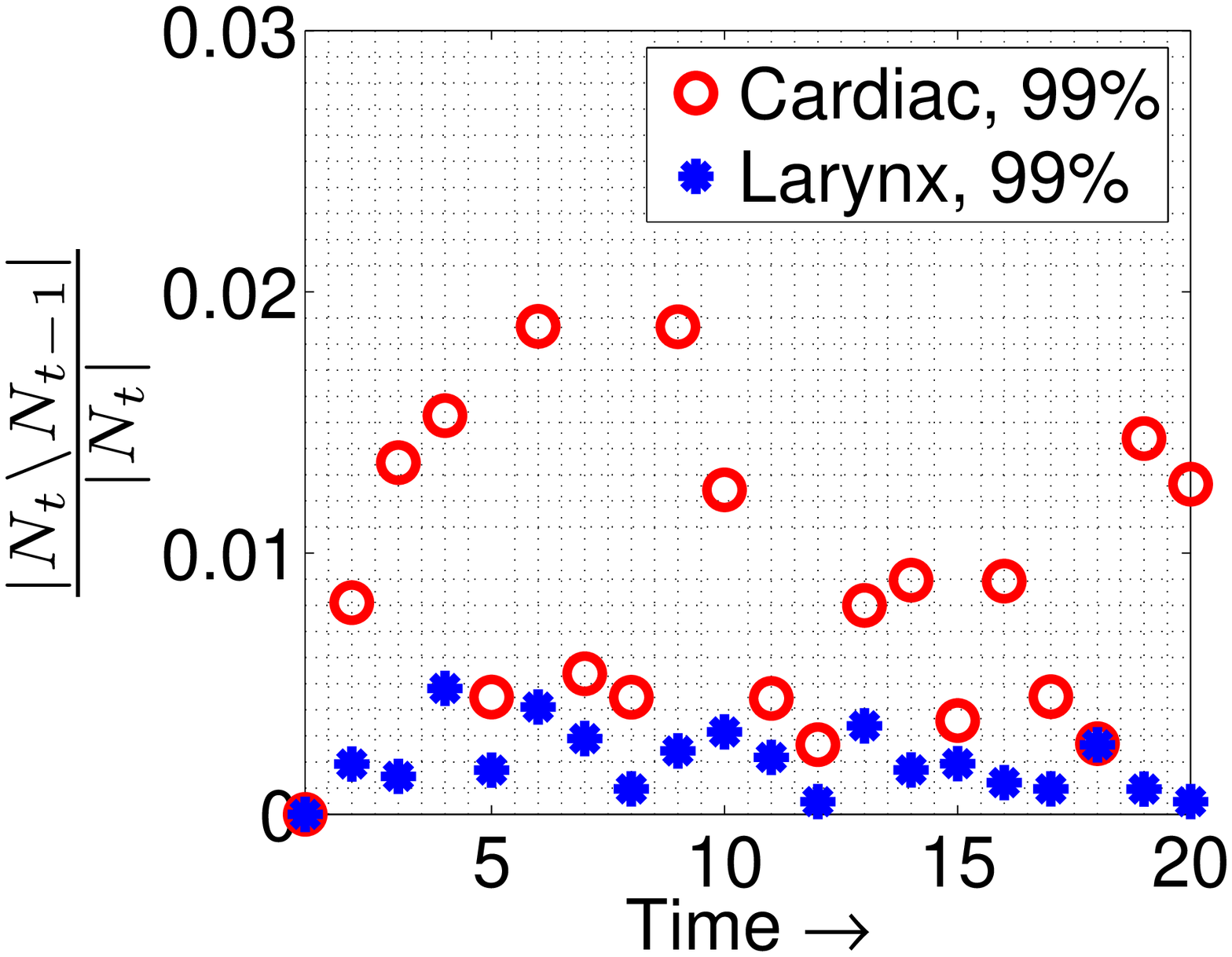,height=2.5cm, width = 4cm} & 
\epsfig{file = 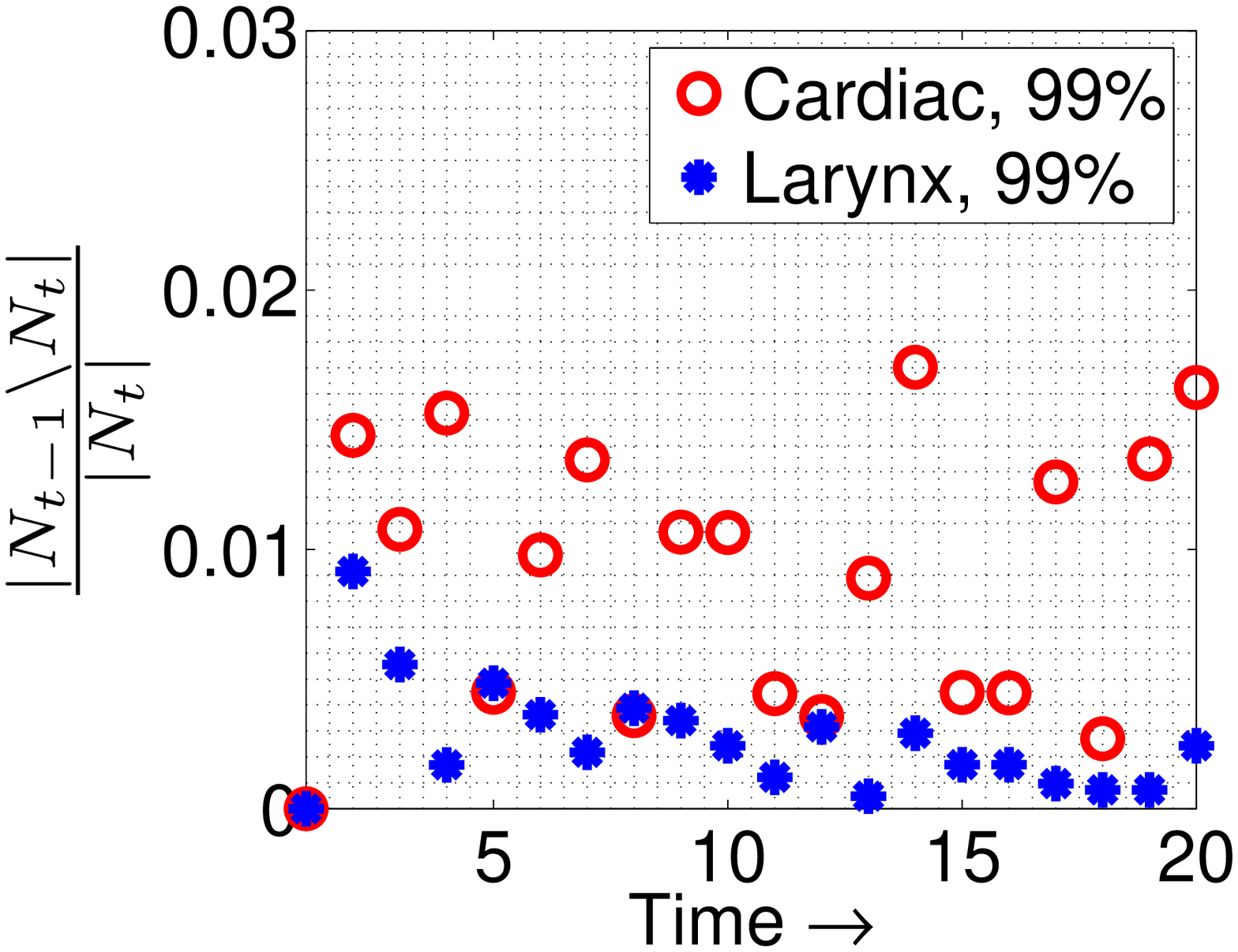,height=2.5cm, width = 4cm}
\end{tabular}
}
}
\vspace{-0.1in}
\caption{\small{
In Fig. \ref{examples}, we show two medical image sequences. In Fig. \ref{supp90_99}, $N_t$ refers to the 99\% energy support of the two-level Daubechies-4 2D discrete wavelet transform (DWT) of these sequences. $|N_t|$ varied between 4121-4183 ($\approx 0.07m$) for larynx and between 1108-1127 ($\approx 0.06m$) for cardiac. We plot the number of additions (left) and the number of removals (right) as a fraction of $|N_t|$. {\em Notice that all changes are less than 2\% of the support size.}
}}
\vspace{-0.2in}
\label{suppchange}
\end{figure}

\Section{Introduction} 
\label{intro}
Consider the problem of recursively and causally reconstructing time sequences of spatially  sparse signals (with unknown and time-varying sparsity patterns) from a limited number of linear incoherent measurements with additive noise. The signals are sparse in some transform domain referred to as the ``sparsity basis" \cite{decodinglp}. An important example is dynamic magnetic resonance (MR) image reconstruction of deforming brain shapes in real-time applications such as MR image guided surgery or functional MRI \cite{interventionalMR}. Human organ images are piecewise smooth and so the wavelet transform is a valid sparsity basis \cite{sparseMRI}. MRI captures a limited number of Fourier coefficients of the image, which are incoherent with respect to the wavelet transform \cite{sparseMRI,dantzig}. Other examples include real-time estimation of time-varying spatial fields using sensor networks, real-time single-pixel video imaging \cite{singlepixel}, and video compression. Due to strong temporal dependency in the signal sequence, it is usually valid to assume that its {\em sparsity pattern (support of the sparsity transform vector) changes slowly over time}. This in  verified in Fig. \ref{suppchange} and in \cite{kfcsmri,chenlu_tip}.

The solution to the static version of the above problem is provided by compressed sensing (CS) \cite{decodinglp,donoho}. CS for noisy observations, e.g. Dantzig selector  \cite{dantzig}, Basis Pursuit Denoising (BPDN) \cite{bpdn,tropp} or Lasso \cite{candesnoise,candes_rip}, have been shown to have small error as long as incoherence assumptions hold.  Most existing solutions for the dynamic problem, e.g. \cite{singlepixel,sparsedynamicMRI}, are non-causal and batch solutions. Batch solutions process the entire time sequence in one go and thus have much higher reconstruction complexity. An alternative would be to apply CS at each time separately (simple CS), which is online and low-complexity, but since it does not use past observations, its reconstruction error is much larger when the number of available observations is small [see Table \ref{static_table} or Figs. \ref{kfcs_better_59}, \ref{dynmri}].%

The question is: for a time sequence of sparse signals, how can we obtain a recursive solution that improves the accuracy of simple CS by using past observations? By {\em ``recursive", we mean a solution that uses only the previous signal estimate and the current observation vector at the current time.} 
The key idea of our proposed solution, LS-CS-residual (LS-CS), is to replace CS on the observation by CS on the least squares (LS) residual computed using the previous support estimate. Its complexity is equal to that of simple CS which is $O(Nm^3)$ where $m$ is the signal length and $N$ is the time duration \cite[Table 1]{ldpc_cs}. Compare this to $O(N^3m^3)$ for a batch solution.%

Other somewhat related work includes \cite{rozell,romberg} (use the previous estimate to speed up the current optimization, but not to improve reconstruction error), and \cite{giannakis} (does not allow the support to change over time). 
Both \cite{romberg}, \cite{giannakis} appeared after \cite{kfcsicip}. The work of \cite{jung_etal} gives an approximate batch solution for dynamic MRI which is quite fast (but offline). Some other related work, but all for reconstructing a single sparse signal, includes \cite{imagereconcs} (uses a recursive algorithm) and \cite{distCS} (related model, but offline algorithm). Finally, none of these bound the reconstruction error or show stability over time.

 In this work, we do ``CS", whether in simple CS or in CS-residual, using the Dantzig selector (DS) \cite{dantzig}. This choice was motivated by the fact that its guarantees are stronger and its results are simpler to apply/modify (they depend only on signal support size) as compared to those for BPDN given in \cite{tropp} (these depend on the actual support elements). 
In later work \cite{kfcsmri,chenlu_tip}, for practical experiments with larger sized images, we have also used BPDN since it runs faster.
Between DS and Lasso ($\ell_2$ constrained $\ell_1$ minimization) \cite{candesnoise,candes_rip}, either can be used \footnote{When we started this work, \cite{candes_rip} had not appeared and the best bounds for Lasso were from \cite{candesnoise}. These are proved under slightly stronger sufficient conditions than the DS results \cite{dantzig}. Also, as argued in the discussion of \cite{dantzig}, the DS bounds are smaller.  Hence we chose to use DS.}. If Lasso is used, by using the results from \cite{candes_rip} as the starting point, results analogous to our Theorems \ref{thm1} and \ref{stabres} can be proved in exactly the same way and the implications will also remain the same.
When using BPDN, obtaining a result similar to Theorem \ref{thm1} is easy \cite{chenlu_tip}. With a little more work it should be possible to also show stability as in Theorem \ref{stabres}.

The LS-CS-residual (LS-CS) algorithm is developed in Sec. \ref{lscs}. We bound its error and compare it with CS in Sec. \ref{csresbound}. Conditions for ``stability" are obtained in Sec. \ref{stability}. Numerical experiments are given in Sec. \ref{sims} and conclusions  in Sec. \ref{conclusions}. Sections marked with ** may be skipped.

\Subsection{Notation}
\label{notation}

The set operations $\cup$, $\cap$, and $\setminus$ have the usual meanings. $T^c$ denotes the complement of $T$ w.r.t. $[1,m]:=[1,2,\dots m]$, i.e. $T^c := [1,m] \setminus T$. $|T|$ denotes the size (cardinality) of $T$.

For a vector, $v$, and a set, $T$, $v_T$ denotes the $|T|$ length sub-vector containing the elements of $v$ corresponding to the indices in the set $T$. $\|v\|_k$ denotes the $\ell_k$ norm of a vector $v$. If just $\|v\|$ is used, it refers to $\|v\|_2$. Also, $v_{(i)}$ refers to the $i^{th}$ largest {\em magnitude} element of $v$ (notation taken from \cite{decodinglp}). Thus, for an $m$ length vector, $|v_{(1)}| \ge |v_{(2)}| \dots \ge |v_{(m)}|$.%

We use the notation $v(S)$ to denote the sub-vector of $v$ containing the $S$ smallest magnitude elements of $v$.

For a matrix $M$, $\|M\|_k$ denotes its induced $k$-norm, while just $\|M\|$ refers to $\|M\|_2$. $M'$ denotes the transpose of $M$. For a tall matrix, $M$, $M^\dag : = (M'M)^{-1}M'$.

For a fat matrix $A$, $A_T$ denotes the sub-matrix obtained by extracting the columns of $A$ corresponding to the indices in $T$.
The $S$-restricted isometry constant \cite{decodinglp}, $\delta_S$, for an $n \times m$ matrix (with $n<m$), $A$, is the smallest real number satisfying
\bea
(1- \delta_S) \|c\|^2 \le \|A_T c\|^2 \le (1 + \delta_S) \|c\|^2
\label{def_delta}
\eea
for all subsets $T \subset [1,m]$ of cardinality $|T| \le S$ and all real vectors $c$ of length $|T|$. 
The restricted orthogonality constant \cite{decodinglp}, $\theta_{S,S'}$, is the smallest real number satisfying
\bea
| {c_1}'{A_{T_1}}'A_{T_2} c_2 | \le \theta_{S,S'} \|c_1\| \|c_2\| 
\label{def_theta}
\eea
for all disjoint sets $T_1, T_2 \subset [1,m]$ with $|T_1| \le S$, $|T_2| \le S'$, $S+S' \le m$, and for all vectors $c_1$, $c_2$ of length $|T_1|$, $|T_2|$.



\Subsection{Problem Definition and Some More Notation}
\label{probdef}
Let $(z_t)_{m \times 1}$ denote the spatial signal at time $t$ and $(y_t)_{n \times 1}$, with $n<m$, denote its noise-corrupted measurements' vector at $t$, i.e. $y_t = H z_t + w_t$ where $w_t$ is measurement noise. The signal, $z_t$, is sparse in a given sparsity basis (e.g. wavelet) with orthonormal basis matrix, $\Phi_{m \times m}$, i.e. $x_t \defn \Phi' z_t$ is a sparse vector. We denote its support by $N_t$. 
Thus the observation model is
\bea
y_t = A x_t + w_t, \ A \defn H \Phi, \ \  \E[w_t]=0, \  \E[w_t w_t']= \sigma^2 I \ \ \ \
\label{obsmod}
\eea
We assume that $A$ has unit norm columns. 
Our goal is to recursively estimate  $x_t$ (or equivalently the signal, $z_t = \Phi x_t$) using $y_1, \dots y_t$. By {\em recursively}, we mean, use only $y_t$ and the estimate from $t-1$, $\xhat_{t-1}$, to compute the estimate at $t$.
%

We state our assumptions after the following definition.%
\bd[Define $\sone$, $\sinf$]
For $A:=H \Phi$, 
\ben
\label{deltas}
\item let $\sone$ denote the largest $S$ for which $\delta_S < 1/2$,
\label{delta3s}
\item let $\sinf$ denote the largest $S$ for which $\delta_{2S}+ \theta_{S,2S} < 1$.
\een
\label{delta_bnd}
\ed
\begin{ass}
In the entire paper, we assume the following.
\ben

\item {\em Sparsity and Slow Support Change. } The support size, $|N_t| \approx |N_0| \ll m$ and the additions, $|N_t \setminus N_{t-1}| \le S_a \ll |N_0|$ and the removals, $|N_{t-1} \setminus N_t| \le S_a \ll |N_0|$.


\item {\em Incoherence. } $A$ satisfies  $S_a < \sinf$ and $|N_t| + k < \sone$ for some $k>0$ (as we argue later $k \ll |N_t|$ suffices).%
\een
\label{initass}
\end{ass}
Sparsity and slow support change is verified in Fig. \ref{suppchange}. Incoherence (approximate orthonormality of $S$-column sub-matrices of $A$) is known to hold with high probability (w.h.p.) when $A$ is a random Gaussian, Rademacher or partial Fourier matrix and $n$ is large enough \cite{decodinglp,dantzig}.

\subsubsection{More Notation}
We use $\xhat_t$ to denote the estimate of $x_t$ given by our algorithm at time $t$ and $\Nhat_t$ to denote its support estimate. To keep notation simple, we avoid using the subscript $t$ wherever possible. We will use the following sets often.%
\bd[Define $T$, $\Delta$, $\Delta_e$]
We use $T  : = \Nhat_{t-1}$ to denote the support estimate from the previous time. This serves as an initial estimate of the current support.
We use $\Delta := N_t \setminus T$ to denote the unknown part of the support at the current time. 
We use $\Delta_e := T \setminus N_t$ to denote the ``erroneous" part of $T$.
\ed
\bd[Define $\tT$, $\tDelta$, $\tDelta_e$]
We use $\tT := \Nhat_t$ to denote the final estimate of the current support.
We use $\tDelta : = N_t \setminus \tT$ to denote the ``misses" in the final estimate and $\tDelta_e := \tT \setminus N_t$ to denote the ``extras". 
\ed

Some more notation - $\xhat_{\CSres}$, $\tDelta_{\dett}$, $\tT_{\dett}$, $\alpha$, $\alpha_{del}$, -  is defined when we give our algorithm in Sec. \ref{lscs}. 


\Section{Least Squares CS-residual (LS-CS)} 
\label{lscs}

Given observation, $y$, the Dantzig selector \cite{dantzig} solves
\bea
\min_{\zeta} \|\zeta\|_1 \ \text{s.t.} \ \|A'(y - A \zeta)\|_\infty < \lambda 
\label{simplecs}
\eea
Now consider the recursive reconstruction problem. If the support of $x_t$, $N_t$, were known at each $t$, we could simply compute its least squares (LS) estimate along $N_t$ while setting all other values to zero. We refer to this estimate as the {\em ``genie-aided" LS estimate}. When $N_t$ is not known, one could do simple CS at each time, i.e. solve (\ref{simplecs}) with $y=y_t$, followed by thresholding the output to estimate its support, and then do the same thing using the support estimate $\Nhat_t$ instead of $N_t$. But in doing so, we are throwing away the information contained in past observations. If the available number of measurements, $n$, is small, this incurs large error [see Table \ref{static_table}].

To use the information contained in past observations, along with the knowledge that support changes slowly, we propose the following idea. Assume for a moment that the support has not changed from $t-1$ to $t$. Use $T := \Nhat_{t-1}$ to compute an initial LS estimate and compute the LS residual, i.e. compute
\bea
(\xhat_{t,\text{init}})_{T} \se {A_{T}}^\dag y_t, \ \ (\xhat_{t,\text{init}})_{T^c} = 0 \nn \\
\tty \se  y_t - A \xhat_{t,\text{init}} \nn  
\label{deftty0}
\eea

Notice that the LS residual, $\tty$, can be rewritten as
\bea
\tty  =  A \beta_t + w_t, \  \beta_t:=x_t - \xhat_{t,\text{init}}
\eea
where $\beta_t$ is a $|T \cup \Delta|$-sparse vector with $(\beta_t)_{(T \cup \Delta)^c} = 0$,
\bea
(\beta_t)_T \se (x_t)_T - {A_T}^\dag y_t  = -  {A_{T}}^\dag (w_t + A_\Delta (x_t)_\Delta), \nn \\
(\beta_t)_\Delta \se (x_t)_\Delta - 0 = (x_t)_\Delta
\label{defbeta}
\eea
The above follows because ${A_T}^\dag A_T = I$ and $y_t = A x_t + w_t = A_{N} ( x_t)_{N} + w_t =  A_{N \cap T} (x_t)_{N \cap T} +  A_{N \cap T^c} (x_t)_{N \cap T^c} + w_t = A_T  (x_t)_T + A_\Delta  (x_t)_\Delta +w_t$. Here $N=N_t$. The last equality holds because $\Delta = N \cap T^c$ and $N \cap T \subseteq T$.%

Notice that $\Delta \subseteq (N_t \setminus N_{t-1}) \cup (N_{t-1} \setminus T)$ and $\Delta_e \subseteq (N_{t-1} \setminus N_t) \cup (T \setminus N_{t-1})$. From Sec. \ref{probdef}, $|N_t \setminus N_{t-1}| \le S_a$ and $|N_{t-1} \setminus N_t| \le S_a$. If $|\Delta|$ and  $|\Delta_e|$ are small enough  (i.e. if $S_a$ is small enough and $T$ is an accurate enough estimate of $N_{t-1}$) and $A$ is incoherent enough to ensure that $\|{A_{T}}'A_\Delta\|$ is small enough, $\beta_t$ will be small (compressible) along $T$. In other words, $\beta_t$ will be only $|\Delta|$-approximately-sparse. In this case, doing CS on $\tty$ should incur much less error than doing CS on $y_t$ (simple CS), which needs to reconstruct a $|N_t|$-sparse signal, $x_t$. This is the key idea of our approach.

We do CS on the LS residual (CS-residual), i.e. we solve (\ref{simplecs}) with $y= \tty$ and denote its output by $\betahat_t$. Now,
\bea
\xhat_{t,\CSres} : = \betahat_t + \xhat_{t,\text{init}}
\label{xhatcsres}
\eea
can serve as one possible estimate of $x_t$. But, as explained in \cite{dantzig}, since  $\betahat_t$ is obtained after $\ell_1$ norm minimization, it will be biased towards zero. Thus, $\xhat_{t,\CSres}$ will also be biased. We can use the Gauss-Dantzig selector trick of \cite{dantzig} to reduce the bias. To do that, we first detect the new additions as follows:
\bea
\Nhat_{t,\dett} \se T \cup \{ i \in [1,m] : |(\xhat_{t,\CSres})_i| > \alpha \}
\label{detectstep}
\eea
and then we use $\tT_{\dett} := \Nhat_{t,\dett}$ to compute an LS estimate
\bea
(\xhat_{t,\dett})_{\tT_{\dett}} \se {A_{\tT_{\dett}}}^\dag  y_t, \ \ (\xhat_{t,\dett})_{(\tT_{\dett})^c} = 0  
\label{secls}
\eea
If $\tT_{\dett}=N_t$, $\xhat_{t,\dett}$ will be unbiased. In fact, it will be the best linear unbiased estimate, in terms of minimizing the mean squared error (MSE). But even if $\tT_{\dett}$ is roughly accurate, the bias and MSE will be significantly reduced.

If the addition threshold, $\alpha$, is not large enough, occasionally there will be some false detections (coefficients whose true value is zero but they wrongly get detected due to error in the CS-residual step). Also, there may have been actual removals from the true support.  This necessitates a ``deletion" step to delete these elements from the support estimate. If this is not done, the estimated support size could keep increasing over time, causing ${A_{T}}'A_{T}$ to become more and more ill-conditioned and the initial LS estimates to become more and more inaccurate. An example of this is shown in Fig. \ref{fast_adds_dels}.

A simple approach to do deletion is to apply thresholding to $\xhat_{t,\dett}$, i.e. to compute
\bea
\Nhat_t \se  \tT_{\dett} \setminus \{ i \in  \tT_{\dett} : |(\xhat_{t,\dett})_i| \le \alpha_{del} \}
\label{delstep}
\eea
The above is better than deleting using $\xhat_{t,\CSres}$ which, as explained above, usually has a larger MSE. 

A final LS estimate can be computed using $\tT:=\Nhat_t$.
\bea
(\xhat_{t})_{\tT} \se {A_{\tT}}^\dag  y_t, \ \ (\xhat_{t})_{\tT^c} = 0  
\label{finalls}
\eea


In most places in the paper, we use {\em ``addition" (``removal")} to refer to additions (removals) from the actual support, while using  {\em``detection" (``deletion")}  to refer to additions (removals) from the support estimate. Occasionally, this is not followed.


%

\Subsection{LS CS-residual (LS-CS) Algorithm and More Notation}
\label{lscsalgo}
We give the LS CS-residual (LS-CS) algorithm below.

\noindent {\em Initialization ($t=0$): } At the initial time, $t=0$, we run simple CS with a large enough number of measurements, $n_0 > n$ (usually much larger), i.e. we solve (\ref{simplecs}) with $y=y_0$ and $A = A_0=H_0 \Phi$ ($H_0$, and hence $A_0$, will be an $n_0 \times m$ matrix). This is followed by support estimation and then LS estimation as in the Gauss-Dantzig selector \cite{dantzig}. We denote the final output by $\xhat_0$ and its estimated support by $\Nhat_0$.
For $t>0$ do,
\ben
\item {\em Initial LS. } Use $T:=\Nhat_{t-1}$ to compute the initial LS estimate, $\xhat_{t,\text{init}}$, and the LS residual, $\tty$, using (\ref{deftty0}).
\label{initls}

\item {\em CS-residual. } Do CS (Dantzig selector) on the LS residual, i.e.  solve (\ref{simplecs}) with $y= \tty$ and denote its output by $\betahat_t$. Compute $\xhat_{t,\CSres}$ using (\ref{xhatcsres}).

\item {\em Detection and LS. } Use (\ref{detectstep}) to detect additions to the support to get $\tT_{\dett}:=\Nhat_{t,\dett}$.
Compute the LS estimate, $\xhat_{t,\dett}$, using $\tT_{\dett}$, as given in (\ref{secls}).
\label{addls}

\item {\em Deletion and LS. }  Use (\ref{delstep}) to detect deletions from the support to get $\tT:=\Nhat_t$.
 Compute the LS estimate, $\xhat_t$, using $\tT$, as given in (\ref{finalls}).
 \label{delete}

\item Output $\xhat_t$ and $\hat{z}_t = \Phi \xhat_t$.  Feedback $\Nhat_t$.
\een
Increment $t$ and go to step \ref{initls}.

We define the following which will be used in Sec. \ref{stability}.
\bd[Define $\tT_{\dett}$, $\tDelta_{\dett}$, $\tDelta_{e,\dett}$]
We define $\tT_{\dett} := \Nhat_{t,\dett}$, $\tDelta_{\dett}:= N_t \setminus \tT_{\dett}$ and $\tDelta_{e,\dett}:= \tT_{\dett} \setminus N_t$.
\ed

%

\Subsection{Threshold Setting Heuristics}
If $n$ is large enough (to ensure that $A_{N_t}$ is well-conditioned), a thumb rule from literature is to set $\alpha$ at roughly the noise level \cite{dantzig}. If the SNR is high enough, we recommend setting $\alpha_{del}$ to a larger value (since in that case, $\xhat_\dett$ will have much smaller error than $\xhat_\CSres$), while in other cases, $\alpha_{del} = \alpha$ is better. Higher $\alpha_{del}$ ensures quicker deletion of extras.

When $n$ is not large enough, instead of setting an explicit threshold $\alpha$, one should keep adding the largest magnitude elements of $\xhat_\dett$ until $A_{\tT_\dett}$ exceeds a condition number threshold. $\alpha_{del}$ can be set to a fraction of the minimum nonzero coefficient value (if known). Also, $\alpha_{del}$ should again be larger than the implicit $\alpha$ being used.%

Another heuristic, which ensures robustness to occasional large noise, is to limit the maximum number of detections at a given time to a little more than $S_a$ (if $S_a$ is known).

\subsection{Kalman filtered CS-residual (KF-CS)}
\label{kfcs}
Now, LS-CS does not use the values of $(\xhat_{t-1})_T$ to improve the current estimate. But, often, in practice, coefficient values also change slowly. To use this information, we can replace the initial LS estimate by a regularized LS estimate. If training data is available to learn a linear prior model for signal coefficients' change, this can be done by using a Kalman filter (KF). 
%
We develop and study the KF-CS algorithm in \cite{kfcsicip,just_kfcs}. As we demonstrate in \cite{just_kfcs}, KF-CS significantly improves upon LS-CS when $n$ is small and so $A_T$ can occasionally become ill-conditioned (this results in LS-CS instability, but does not affect KF-CS much, as long as this occurs only occasionally).%



\Section{Bounding CS-residual Error}
\label{csresbound}

We first bound the CS-residual reconstruction error and compare it with the bound on CS error. In Sec. \ref{tightbnd}, we give a tighter bound on the CS-residual error, but which holds under stronger assumptions. All bounds depend on $|T|$, $|\Delta|$.

{\em To simplify notation, in this section, we remove the subscript $t$}. Consider reconstructing $x$ with support, $N$, from $y:=Ax + w$. The support can be written as $N = (T \cup \Delta) \setminus \Delta_e$ where $T$ is the ``known" part of the support (equal to support estimate from the previous time), $\Delta_e := T \setminus N$ and $\Delta:= N \setminus T$.

\Subsection{Bounding CS-residual Error}
\label{csresbound1}
If $n$ is large enough so that $|T|+|\Delta|=|N|+|\Delta_e| \le \sinf$, then we can use the bounds given in Theorems 1.1 or 1.2 of \cite{dantzig} to bound the CS-residual error. But recall that CS-residual is primarily designed for situations where $n$ is smaller. It applies CS to the observation residual $\tilde{y} = A \beta + w$ where $\beta := x - \xhat_{\text{init}}$ is a $(|T|+|\Delta|)$-sparse signal, that is compressible (small) along $T$. To bound its error, we first prove Lemma \ref{thm13} which modifies Theorem 1.3 of \cite{dantzig} to apply it to ``sparse-compressible signals", i.e. sparse signals that are partly (or fully) compressible. Next, we bound $\|\beta_T\|$ (the ``compressible" part of $\beta$). Finally we use this lemma along with the bound on $\|\beta_T\|$ to obtain the CS-residual error bound.

\begin{lemma}[CS error bound - sparse-compressible signal]
Assume that $\|w\|_\infty \le \frac{\lambda}{\|A\|_1}$ (bounded noise). Let $\zeta$ is an $S_{nz}$-sparse vector with support $T_{nz}$, and we measure $y:=A\zeta+w$. Its estimate, $\zetahat$, obtained by solving (\ref{simplecs}), obeys the following.
For all sets $\rest \subseteq T_{nz}$ of size $|\rest|= S_{nz}-S$ and for all $1 \le S \le \min(\sinf,S_{nz})$,
\bea
\label{ell1term}
\|\zeta - \zetahat\|^2 \sle C_2(S) S  \lambda^2 + C_3(S) \frac{\| (\zeta)_{{\rest}} \|_1^2}{S} \nn  \\
\label{ell2term}
\sle C_2(S) S  \lambda^2 + C_3(S) \frac{(S_{nz} - S)}{S} \| (\zeta)_{{\rest}} \|^2,   \nn \\
 \text{where~~} C_2(S) \sdefn \frac{48}{(1-\delta_{2S} - \theta_{S,2S})^2},   \nn \\ 
\text{and~~} C_3(S) \sdefn 8 + \frac{24 \theta_{S,2S}^2}{(1-\delta_{2S} - \theta_{S,2S})^2} 
\label{defc2c3}
\eea
\label{thm13}
\end{lemma}
The proof is given in Appendix \ref{thm13_proof}.
Recall that $\sinf$ is defined in Definition \ref{delta_bnd}.

Recall that $\beta = x - \xhat_{\text{init}}$ can be rewritten as
\bea
\beta_T \se   - ({A_T}'A_T)^{-1} {A_T}'(A_\Delta x_\Delta + w)  \nn \\ 
\beta_\Delta \se x_\Delta, \ \ \ (\beta)_{(T \cup \Delta)^c} = 0
\label{defbeta2}
\eea
As long as $\delta_{|T|}< 1$, $\|\beta_T\|^2$ can be bounded as follows.
\bea
\|\beta_T\|^2 \sle  2  [ \|({A_T}'A_T)^{-1} {A_T}'A_\Delta x_\Delta\|^2 + \|({A_T}'A_T)^{-1} {A_T}' w \|^2 ]   \nn \\
\sle  2 \frac{{\theta_{|T|,|\Delta|}}^2}{(1-\delta_{|T|})^2} \|x_\Delta\|^2 + \frac{2}{(1-\delta_{|T|})} \|w\|^2
\label{betabound}
\eea
The above follows by using (a) $\|({A_T}'A_T)^{-1}{A_T}'\|^2 = \|({A_T}'A_T)^{-1}\| \le 1/(1-\delta_{|T|})$ (if $\delta_{|T|}<1$) and (b) $\|{A_T}'A_\Delta\| \le \theta_{|T|,|\Delta|}$. (a) follows because $\|({A_T}'A_T)^{-1}\|$ is the inverse of the minimum eigenvalue of ${A_T}'A_T$. (b) follows from (\ref{def_theta}) by setting $c_1 = {A_{T_1}}'A_{T_2} c_2$, $T_1 =T$, and $T_2=\Delta$.%

From (\ref{betabound}), it is clear that if the noise is small and if $|\Delta|$, $|\Delta_e|$ are small enough so that $\theta$ is small, $\|\beta_T\|$ is small. Using (\ref{betabound}) along with Lemma \ref{thm13}, we can prove the following.

\begin{theorem}[CS-residual error bound]
Assume that $|T|:=|N|+|\Delta_e|-|\Delta| \le \sone$ and $\|w\|_\infty \le \frac{\lambda}{\|A\|_1}$. Then,
\bea
&& \| x -\xhat_{\CSres} \|^2 \le  \min_{1 \le S \le \min(\sinf,|T|+|\Delta|)} F_{\CSres}(S), \ \text{where}    \nn \\
&& F_{\CSres}(S):= [ C_2(S) S \lambda^2 + C_3(S) \frac{|T|+|\Delta|-S}{S} B(S) ], \nn \\
&& B(S)=\left\{ \begin{array}{ll}
8{\theta}^2 \|x_\Delta\|^2 +  4 \|w\|^2  &  \text{if} \ S \ge |\Delta| \nn \\
8{\theta}^2 \|x_\Delta\|^2 +  4 \|w\|^2 + \|x_\Delta(|\Delta|-S)\|^2  &  \text{if} \  S < |\Delta|
\end{array} \right. \nn \\
\label{tighterbnd}
\eea
where $\theta := \theta_{|T|,|\Delta|}$ and $C_2(S)$, $C_3(S)$ are defined in (\ref{defc2c3}).
\label{thm1}
\end{theorem}

The proof is given in Appendix \ref{thm13_proof}. {\em Recall: $x_\Delta(K)$ is the vector containing the $K$ smallest magnitude elements of $x_\Delta$.}

A simple corollary of the above result follows by applying it for a particular value of $S$, $S=|\Delta|$ when $|\Delta| > 0$. This will usually result in the smallest bound in situations where $\sinf$ is not much larger than $|\Delta|$. 
When $|\Delta|=0$, $\beta_T = {A_T}^\dag w$ which is anyway quite small. It is not immediately clear which value of $S$ is the best. We retain the $\min$ in this case. We also bound $\|w\|$ by its maximum value, $\sqrt{n}\lambda/\|A\|_1$.
\begin{corollary}[simpler CS-residual error bound]
Assume that  $\|w\|_\infty \le {\lambda}/{\|A\|_1}$, $|\Delta| \le \sinf$ and $|T| \le \sone$.
\\
1) {If $|\Delta|>0$}, 
\bea
&& \|x - \xhat_{\CSres}\|^2 \le C' + C''{\theta_{|T|,|\Delta|}}^2 \|x_\Delta\|^2, \ \ \text{where} \nn \\
&& C' \equiv C'(|T|,|\Delta|) := C_2(|\Delta|) |\Delta| \lambda^2 + 4 C_3(|\Delta|) \frac{|T|}{|\Delta|} \frac{n\lambda^2}{\|A\|_1^2} \nn \\
&& C'' \equiv C''(|T|,|\Delta|) := 8 C_3(|\Delta|) |T| 
\label{defcprime}
\eea
2) {If $|\Delta|=0$}, $\|x - \xhat_{\CSres}\|^2 \le B_0$ where
\bea
B_{0} := \min_{1 \le S \le \sinf} [ C_2(S) S \lambda^2 + C_3(S) \frac{|T|-S}{S} \frac{4n\lambda^2}{\|A\|_1^2} ] \ \ \ \ \
\label{defB0}
\eea
where $C_2(S)$, $C_3(S)$ are defined in (\ref{defc2c3}).
\label{cor1}
\end{corollary}
This corollary is used in the LS-CS stability result.%


\begin{remark}
It is easy to modify the above results for {\em Gaussian noise}, $w \sim \n(0,\sigma^2I)$, in a fashion analogous to the results of \cite{dantzig}. Like  \cite{dantzig}, we will also get ``large probability" results. We do not do this here because any large probability result will make the study of stability over time difficult.
\end{remark}

\begin{remark}
In the bounds of Theorem \ref{thm1} or Corollary \ref{cor1}, there is a term that is proportional to $(|T|+|\Delta|-S)$ or to $|T|$ respectively. This comes from Lemma \ref{thm13} when we bound the $\ell_1$ norm term, $\| (\zeta)_{{\rest}} \|_1^2$, by $(S_{nz}-S)$ times $\| (\zeta)_{{\rest}} \|^2$. A similar term is also there in the CS bound given below in (\ref{csbnd}). This is the bound for CS error which holds under the same weak assumptions as those used by our result\footnote{A term containing the $\ell_1$ norm of the ``compressible" part appears in all bounds for CS for compressible signals, e.g. \cite[Thm 1.3]{dantzig}, and hence also appears when we try to bound CS error for sparse signals with not-enough measurements.}.
\end{remark}

\Subsection{Comparing CS-residual and CS error bounds}
\label{comparecs}
We now compare the CS-residual bound with that of CS.
\begin{remark}
By showing that the upper bound on CS-residual error is much smaller, we only show that the performance guarantees for CS-residual are better than those for CS. To actually compare their errors, we use simulations.
\end{remark}

To compare the CS-residual bound with CS, first note that the CS error bounds for sparse signals given in Theorems 1.1 and 1.2 of \cite{dantzig} apply  only when $\delta_{2S} + \theta_{S,2S} < 1$ holds for $S=|N|$, i.e. when $|N| \le \sinf$. When $n$ is small and this does not hold, these results are not applicable. On the other hand, Lemma \ref{thm13} does not assume anything about $\sinf$. 
Let $\xhat_{CS}$ denote the simple CS output. Using Lemma \ref{thm13}, it is easy to see that 
\bea 
&& \|x - \xhat_{CS}\|^2 \le \min_{1 \le S \le \min(\sinf,|N|)} F_{CS}(S), \ \text{where} \nn \\
&& F_{CS}(S):=[ C_2(S) S \lambda^2 + C_3(S) \frac{|N|-S}{S} B_{CS}(S) ], \nn \\
&& B_{CS}(S) := \|x_N(|N|-S)\|^2
\label{csbnd}
\eea

Compare (\ref{tighterbnd}) with (\ref{csbnd}) under the following assumptions.
\ben
\item  The magnitude of the largest element of $x_\Delta$ is smaller than or equal to that of the smallest element of $x_{N \setminus \Delta}$, i.e. $|(x_\Delta)_{(1)}| \le |(x_{N\setminus \Delta})_{(N\setminus \Delta)}|$.
 This is reasonable since $\Delta$ contains the recently added elements which will typically be smaller while ${N \setminus \Delta}$ contains the previously added elements which should have a larger value.
\label{Delsmall}

\item $|\Delta|$, $|\Delta_e|$ are small enough (i.e. $S_a$ is small enough and $T \approx N_{t-1}$) 
and the noise is small enough so that
\ben
\item $|\Delta| \le 0.1|N|$ and $|\Delta_e| \le  0.1|N|$ (this ensures that $|T| \le 1.1|N|$).%
\label{smallerr}

\item ${\theta_{|T|,|\Delta|}}^2 < 1/8$, and
 $\|w\|^2 \le \frac{\|x_{N\setminus \Delta}(|\Delta|)\|^2(1-8 \theta^2)}{4}$.
\label{noisesmall}
\een
\item $n$ is small so that $\sinf =0.2|N|$, but is just large enough so that $\sone \ge 1.1|N|$. $\sone \ge 1.1|N|$ along with assumption \ref{smallerr} ensures that $\delta_{|T|} < 1/2$.
\een
The above assumptions ensure that $\|\beta_T\|^2 \le 8{\theta}^2 \|x_\Delta\|^2 +  4 \|w\|^2 \le 8{\theta}^2  \|x_{N \setminus \Delta}(|\Delta|)\|^2 +  4 \|w\|^2 \le \|x_{N \setminus \Delta}(|\Delta|)\|^2$, i.e.  $\beta_T$ is ``compressible enough".

Under the above assumptions, we show that $F_{\CSres}(S)$ in (\ref{tighterbnd}) is significantly smaller than $F_{CS}(S)$ in (\ref{csbnd}) for each value of $S$ and hence the same will hold for the upper bounds\footnote{\label{minS}Notice that $\frac{F_{\CSres}(S)}{F_{CS}(S)} \le a$ for all $S$ implies that $\frac{\min_{S} F_{\CSres}(S)}{F_{CS}(S)} \le a$ for all $S$. Since this holds for all $S$, it also holds for the max taken over $S$, i.e. $\frac{\min_{S} F_{\CSres}(S)}{\min_{S}F_{CS}(S)} = \max_{S} \frac{\min_{S} F_{\CSres}(S)}{F_{CS}(S)} \le a$.}.



For any $S$, the first term in $F_{\CSres}(S)$ and $F_{CS}(S)$ is the same. In the second term, the main difference is in $B(S)$ versus $B_{CS}(S)$. The constants are almost the same, their ratio is $\frac{|N|+|\Delta_e|-S}{|N|-S} = 1+ \frac{|\Delta_e|}{|N|-S} \le 9/8$ (follows since $S \le \sinf=0.2|N|$ and $|\Delta_e| \le 0.1|N|$). Thus if we can show that $B(S)$ is much smaller than $B_{CS}(S)$, we will be done.%

First consider  $|\Delta| \le S \le 0.2|N|$. In this case, 
\bea
B(S) \se 8{\theta}^2 \|x_\Delta\|^2 +  4 \|w\|^2  \nn \\
\sle  \|x_{N\setminus \Delta}(|\Delta|)\|^2 \le \|x_{N\setminus \Delta}(0.1|N|)\|^2 \nn \\
B_{CS}(S) \ge B_{CS}(0.2|N|) \se \|x_N(0.8|N|)\|^2 \nn \\
\se  \|x_\Delta\|^2 + \|x_{N \setminus \Delta} (0.8|N|-|\Delta|)\|^2 \nn \\
& > &  0 +  \|x_{N \setminus \Delta} (0.7|N|)\|^2 \nn \\
\sge 7 \|x_{N \setminus \Delta} (0.1|N|)\|^2 \ge  7 B(S)
\eea

Now consider $1 \le S <  |\Delta|$  
\bea
B(S) \se 8{\theta}^2 \|x_\Delta\|^2 +  4 \|w\|^2  + \|x_\Delta(|\Delta|-S)\|^2    \nn \\
\sle \|x_{N\setminus \Delta}(|\Delta|)\|^2  + \|x_\Delta\|^2  \nn \\
\sle 2\|x_{N\setminus \Delta}(|\Delta|)\|^2 \le  2\|x_{N\setminus \Delta}(0.1|N|)\|^2 \nn \\
B_{CS}(S) > B_{CS}(|\Delta|) \sge B_{CS}(0.1|N|) = \|x_N(0.9|N|)\|^2 \nn \\
 \sge  \|x_{N \setminus \Delta} (0.8|N|)\|^2  \nn \\
\sge 8 \|x_{N \setminus \Delta} (0.1|N|)\|^2  \ge 4 B(S)
\eea
Thus $\frac{B(S)}{B_{CS}(S)} \le 1/4$ in all cases. 
 Denote the common first term in $F_\CSres$ and $F_{CS}$  by $\text{T1}$. Denote the second terms by $\text{T2}_\CSres$ and $\text{T2}_{CS}$. 
Thus, $\frac{\text{T2}_\CSres}{\text{T2}_{CS}} =  \frac{B(S)}{B_{CS}(S)} \frac{|N|+|\Delta_e|-S}{|N|-S}  \le (1/4) (9/8)=9/32$.  Thus, $F_{\CSres}(S) \le F_{CS}(S)$ for all $S \le \sinf$. By footnote \ref{minS} the same holds for the bounds.

Furthermore, if the noise is small enough, and for $S \le \sinf$, the second term is the dominant term in $F_{CS}(S)$, i.e. $\text{T1} \ll \text{T2}_{CS}$. Then $\frac{F_{\CSres}(S)}{F_{CS}(S)} = \frac{\text{T1}}{\text{T1}+\text{T2}_{CS}} + \frac{\text{T2}_{\CSres}}{\text{T1}+\text{T2}_{CS}} \approx \frac{\text{T2}_\CSres}{\text{T2}_{CS}} \le 9/32$, i.e. $F_{\CSres}(S)/F_{CS}(S)$ is also roughly less than $9/32$ for all $S$.
From footnote \ref{minS}, this means that {\em the CS-residual bound is also roughly $(9/32)$ times the CS bound (is significantly smaller).}

If $|\Delta|=0$, assumption \ref{noisesmall} does not hold and so the above comparison does not hold. But clearly, under high enough SNR, $B_0$ in Corollary \ref{cor1} is much smaller than (\ref{csbnd}).%

\subsubsection{Monte Carlo Comparison}
We compared CS-residual error with CS error using simulations for a single time instant problem with $m=200$, $|N|=20$, $|\Delta|=|\Delta_e|=0.1|N|=2$ and for $n=45,59,100$. 
We compare the normalized MSE's in Table \ref{static_table}. The CS-residual error is much smaller than the CS error except when $n=100$ (large) in which case the errors are roughly equal. See Sec. \ref{staticsim} for details.

\Subsection{Tighter CS-residual Bound under Stronger Assumptions**}
\label{tightbnd}
To address an anonymous reviewer's comment, we give below a tighter error bound for CS-residual (does not contain a term proportional to $|T|$). But this also holds under a stronger assumption. This section can be skipped in a quick reading.

 Using (\ref{defbeta2}), $\| (\beta)_{T} \|_1 \le \|{A_T}^\dag A_\Delta\|_1 \|x_\Delta\|_1 + \|{A_T}^\dag w\|_1$. Notice that if the noise is small and $|\Delta|$ is small, $\| (\beta)_{T} \|_1$ will be small. In particular, if $\| (\beta)_{T} \|_1 \le b \| x_\Delta \|_1$, by applying the first inequality of Lemma \ref{thm13} with $\zeta = \beta$, $S_{nz}=|T|+|\Delta|$, $S = |\Delta|$ and $\rest=T$; using $\| x_\Delta \|_1\le \sqrt{|\Delta|} \| x_\Delta \|$; and combining the resulting bound with that given in Corollary \ref{cor1}, we get the following.%
\begin{corollary}
Assume that
\ben
\item $\|w\|_\infty \le \frac{\lambda}{\|A\|_1}$, $|\Delta| \le \sinf$, $|T| \le \sone$,
\item  $\|{A_T}^\dag A_\Delta\|_1 < c$ and $\|x_\Delta\|_1 > \frac{\|{A_T}^\dag w\|_1}{b-\|{A_T}^\dag A_\Delta\|_1}$ for a $b>c$%
\label{x_delta_large}
\een
then,
\bea
&& \|x - \xhat_{\CSres}\|^2 \ \le  \ C_2(|\Delta|) |\Delta|  \lambda^2 \ + \ C_3(|\Delta|) \times   \nn \\
&& \min(b^2 \|x_\Delta\|^2, \ {8|T|}{\theta}^2 \|x_\Delta\|^2 + {4|T|} \frac{n\lambda^2}{\|A\|_1^2})
\label{cor2bnd}
\eea
If $|\Delta|=0$,  condition \ref{x_delta_large} cannot hold. In this case, $\|x - \xhat_{\CSres}\|^2 \le B_0$ with $B_0$ defined in Corollary \ref{cor1}.
$\blacksquare$%
\label{cor2}
\end{corollary}

 Notice that {\em the first term in the $\min$ does not contain ${|T|}$.} The above bound is tighter when this first term is smaller, i.e. $b$ is small enough (happens if  $|\Delta|$ small but $\|x_\Delta\|_\infty$ large).%

\Section{LS-CS Stability}
\label{stability}

So far we bounded CS-residual error as a function of $|T|,|\Delta|$. The bound is small as long as $|\Delta_e|$ and $|\Delta|$ are small. A similar bound on LS-CS error as a function of $|\tT|,|\tDelta|$ is easy to obtain. The next questions are:
\ben
\item Under what conditions on the measurement model and the signal model, will the number of  extras, $|\tDelta_e|$, and the number of misses, $|\tDelta|$, and hence also $|\Delta_e|$, $|\Delta|$, be bounded by a time-invariant value, i.e. be ``stable"? This will imply a time-invariant bound on LS-CS error.
\item If additions/removals occur every-so-often, under what conditions can we claim that $|\tDelta_e|$, $|\tDelta|$ will become zero within a finite delay of an addition time? This will mean that the LS-CS estimate becomes equal to the genie-aided LS estimate (LS estimate computed using $N_t$).
\een
The answers to both these questions are interrelated and are given in a single theorem. Of course as mentioned earlier, ``stability" is meaningful only if the bounds on the misses and extras are small compared to the support size.


\Subsection{Signal Model}
\label{signalmodel}

For studying stability, we need to assume a signal model. We assume the following deterministic model that (a) assumes a nonzero delay between new coefficient addition times, (b) allows a new coefficient magnitude to gradually increase from zero for sometime and finally reach a constant value and (c) allows coefficients to gradually decrease and become zero (get removed from support). At $t=0$, we assume that $x_0$ is $(S_0-S_a)$ sparse with all ``large" coefficients with values $\pm M$.

\begin{sigmodel} The model is as follows.
\ben
\item {\em Initialization. } At $t=0$, $x_0$ is $(S_0-S_a)$ sparse. All its nonzero coefficients have values $\pm M$.

\item {\em Addition. } At $t=t_j =1+(j-1)d$, for all $j \ge 1$, $S_a$ new coefficients get added. Denote the set of indices of coefficients added at $t=t_j$ by $\add = \add(j)$. A new coefficient, $i \in \add$, gets added at an initial magnitude $a_i$ (its sign can be $\pm1$) and then its magnitude increases at a rate $a_i$ until it either reaches $M$ or for $d$ time units. Thus, the maximum magnitude of the $i^{th}$ coefficient is $\min(M,da_i)$ for $i \notin N_0$, and is $M$ for $i \in N_0$.

\item {\em Removal. } $S_a$ coefficients get removed at $t=t_{j+1}-1 = jd$ for all $j \ge1$. Denote the set of indices of coefficients which get removed at $t_{j+1}-1$ by $\rem = \rem(j)$. During  $[t_{j+1}-r, t_{j+1}-1]$, the elements of  $\rem$ start to decrease and become zero at $t=t_{j+1}-1$.  For coefficient, $i$, the rate of decrease is $\min(M,da_i)/r$ per unit time.

\item The sets $\add(j)$ and $\rem(j)$ are disjoint, i.e. the coefficients that just got added do not get removed.
\een
Thus at any $t \in [t_j, t_{j+1}-r-1]$, the support can be split as $\add$ {\em (increasing coefficients)} and $N \setminus \add$ {\em (constant coefficients)}, where $N = N_t = N_{t_j}$. At any $t \in [t_{j+1}-r, t_{j+1}-2]$, it can be split as $\add$ (increasing), $\rem$ {\em (decreasing)}, $N \setminus (\add \cup \rem)$ (constant). At $t=t_{j+1}-1$, $N=N_t = N_{t_j} \setminus \rem$ (all constant).%
\label{sigmod1}
\end{sigmodel}

Notice that in the above model the signal support size remains roughly constant. It is $S_0$ or $(S_0-S_a)$ at all times. Also, the maximum signal power is bounded by $S_0M^2$.

\Subsection{Three Key Lemmas}
\label{keylemmas} 
Proving stability, i.e. showing that the number of misses, $|\tDelta|$, and extras, $|\tDelta_e|$, remain bounded, requires finding sufficient conditions for the following three things to hold at the current time: (a) one, or a certain number of, large undetected coefficients definitely get detected; (b) large enough detected coefficients definitely do not get falsely deleted, and (c) every-so-often the extras (false detects or true removals) definitely do get deleted. (a) and (b) are used to ensure that $|\tDelta|$ remains bounded while (c) is  used to ensure that $|\tDelta_e|$, and hence $|\tT| \le |N|+|\tDelta_e|$, remains bounded. These three things are done in the following three lemmas.

\begin{lemma}[Detection condition]
Assume that  $|T| \le S_T$, $|\Delta| \le S_\Delta$, and  $\|w\|_\infty \le \lambda/\|A\|_1$. 
The current largest magnitude undetected element, $(x_\Delta)_{(1)}$, will definitely get detected at the current time if $S_T \le \sone$, $S_\Delta \le \sinf$,
\bea
&& 2{\theta_{S_T,S_\Delta}}^2 S_\Delta C''(S_T,S_\Delta) < 1, \ \text{and}        \nn \\
&& \max_{|\Delta| \le S_\Delta} \frac{2\alpha^2 + 2C'(S_T,|\Delta|)}{1- 2 \theta_{S_T,|\Delta|}^2 |\Delta| C''(S_T,|\Delta|)}  < (x_\Delta)_{(1)}^2 \ \ \ \ \
\label{largestdet}
\eea
\label{largest}
\end{lemma}

\begin{lemma}[No false deletion condition] 
Assume that $\|w\|_\infty \le \lambda/\|A\|_1$, $|\tT_\dett| \le S_T$ and $|\tDelta_\dett| \le S_\Delta$. For a given $b_1$, let $T_l:=\{i \in \tT_\dett:x_i^2 \ge b_1\}$. All $i \in T_l$ will not get (falsely) deleted at the current time if $S_T \le \sone$, and
\bea
b_1^2 >2 \alpha_{del}^2 + \frac{8n\lambda^2}{\|A\|_1^2} + 16 {\theta_{S_T,S_\Delta}}^2 |\tDelta_\dett| \|x_{\tDelta_\dett}\|_\infty^2 
\eea
\label{nofalsedelscond}
\end{lemma}

\begin{lemma}[Deletion condition]
Assume that $\|w\|_\infty \le \lambda/\|A\|_1$, $|\tT_\dett| \le S_T$ and $|\tDelta_\dett| \le S_\Delta$. All elements of $\tDelta_{e,\dett}$ will get deleted if $S_T \le \sone$ and
\label{truedelscond}
$
\alpha_{del}^2 \ge \frac{4n\lambda^2}{\|A\|_1^2} + 8 {\theta_{S_T,S_\Delta}}^2 |\tDelta_\dett| \|x_{\tDelta_\dett}\|_\infty^2 
$.
\end{lemma}

These lemmas follow easily from Corollary \ref{cor1} and a few simple facts. They are proved in Appendix \ref{append_lscs_stab_1}.

\Subsection{The Main Result}

\begin{figure}[t!]
\centerline{
\epsfig{file = 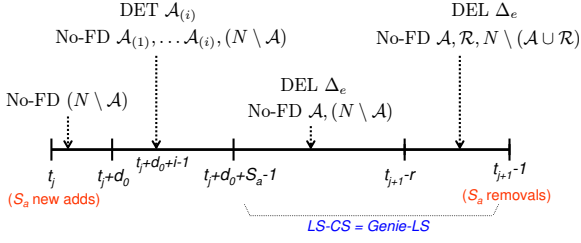, width=8cm}
}
\caption{\small{Our approach to show stability (prove Theorem \ref{stabres}). We split $[t_j,t_{j+1}-1]$ into the four sub-intervals shown above and ensure no false deletion (No-FD) / detection (DET)  / deletion (DEL) of the coefficients listed in each.
The notation $\add_{(i)}$ refers to the $i^{th}$ largest increasing coefficient. Recall: in $[t_j,t_{j+1}-r-1]$, $\add$ is the increasing coefficients' set and $N \setminus \add$ is the constant coefficients' set.%
}}
\vspace{-0.2in}
\label{stabfig}
\end{figure}

We analyze the LS-CS algorithm given in Sec. \ref{lscsalgo}. By running simple CS at $t=0$ with an appropriate number of measurements, $n_0 > n$ (usually much larger), we assume that we detect all nonzero coefficients and there are no false detects, i.e. $\Nhat_0 = N_0$. This assumption is made for simplicity.

For stability, we need to ensure that within a finite delay of a new addition time, all new additions definitely get detected (call this delay the ``detection delay"). This needs to be done while ensuring that there are no false deletions of either the constant or the definitely detected increasing coefficients. Further, (a) by letting the  delay between two addition times be larger than the ``detection delay" plus the coefficient decrease time, $r$, and (b) by setting the deletion threshold high enough to definitely delete the extras in the duration after all detections are done, we can show stability.

To obtain our result, the above is done by splitting $[t_j, t_{j+1}-1]$ into the four subintervals shown in Fig. \ref{stabfig} and using the lemmas from the previous subsection to find sufficient conditions so that the following hold for some $d_0 < d$:
\ben
\item At all $t \in [t_j, t_j+d_0-1]$, there is no false deletion of the constant coefficients (during this time the increasing coefficients may be too small and we do not care if they get detected or not). This ensures that the number of misses do not increase beyond $S_a$.

\item At $t=t_j+d_0+i-1$, for $i=1,\dots S_a$,  (a) the $i^{th}$ largest increasing coefficient definitely gets detected, and (b)  all constant coefficients and the first $i$ largest increasing coefficients do not get falsely deleted. This ensures that by $t=t_j+d_0+S_a-1$, the number of misses becomes zero, i.e. the ``detection delay" is $d_0+S_a-1$.%

\item  At $t=t_j+d_0+S_a-1$,  all false detects get deleted. This is needed to keep $|T|$ bounded.

\item At all $t \in [t_j+d_0+S_a,t_{j+1}-r-1]$,  (a) the current falsely detected set is immediately deleted and (b) none of the constant or increasing coefficients get falsely deleted.

\item At all $t \in [t_{j+1}-r, t_{j+1}-2]$, (a) the current falsely detected set is deleted and (b) none of the decreasing, constant or increasing coefficients are falsely deleted.
\item At $t_{j+1}-1$, all falsely detected and removed coefficients are deleted and there is no false deletion.
\een

Doing the above leads to the following result.

\begin{theorem}[LS-CS Stability]
Under Signal Model \ref{sigmod1}, if there exists a $d_0 < d$, so that the following conditions hold:
\ben
\item {\em (initialization)}  all elements of $x_0$ get correctly detected and there are no false additions, i.e. $\Nhat_0 = N_0$,

\item {\em (algorithm - thresholds)} we set $\alpha_{del} = 2 \sqrt{n} \lambda/\|A\|_1$ and we set $\alpha$ large enough so that there are at most $f$ false detections per unit time,
\label{algo_thresh}

\item {\em (measurement model)}
\label{measmodass}
\ben
\item $\|w\|_\infty \le \lambda/\|A\|_1$, $S_a \le \sinf$, $S_0+f(d_0+S_a) \le \sone$, and
\label{sone_sinf_bnd}
\item $2 {\theta_{S_T,S_\Delta}}^2 S_\Delta C''(S_T,S_\Delta) < 1$ with $S_T=S_0+f(d_0+S_a)$ and $S_\Delta=S_a$
\label{thetabnd}
\een

\item {\em (signal model - additions \& no false deletions of increasing coefficients)}  the following hold for all $i=1, \dots S_a$ and for all $\add = \add(j)$ for all $j$:
\label{additioncond}
\ben
\item with $S_T=S_0+f(d_0+i-1)$ and $S_\Delta=S_a-i+1$,%
\label{add_incr}
\bea
&& \min(M, (d_0+i) (a_\add)_{(i)} )^2 > \nn \\
&& \max_{|\Delta| \le S_\Delta} \frac{2\alpha^2 + 2C'(S_T,|\Delta|)}{1- 2(\theta_{S_T,|\Delta|})^2 |\Delta| C''(S_T,|\Delta|)} \nn
\eea

\item with $S_T=S_0+f(d_0+i)$, $S_\Delta=S_a-i$ and $(a_\add)_{(S_a+1)} \equiv 0$,
\label{nofd_incr}
\bea
&& \min(M, (d_0+i) (a_\add)_{(i)} )^2 >
 2\alpha_{del}^2  + ({8n\lambda^2}/{\|A\|_1^2})  \nn \\
&& + 16{\theta_{S_T,S_\Delta}}^2 (S_a-i) \min(M,(d_0+i) (a_\add)_{(i+1)})^2 \nn
\eea
\een

\item {\em (signal model - no false deletions of constant coefficients)} with $S_T=S_0+f(d_0+S_a)$, $S_\Delta=S_a$,
\label{nofd_const}
\bea
&& \min(M,d \min_i a_i)^2 > 2\alpha_{del}^2  + ({8n\lambda^2}/{\|A\|_1^2})  \nn \\
&& + 16 {\theta_{S_T,S_\Delta}}^2 S_a  \min(M,(d_0+S_a) \max_i a_i)^2 \nn
\eea

\item {\em (signal model - no false deletions of decreasing coeff's)}%
\label{nofd_decr}
\bea
\min(M,d \min_i a_i)^2 > r^2 (2\alpha_{del}^2 +  ({4n\lambda^2}/{\|A\|_1^2})) \nn
\eea

\item {\em (signal model - delay b/w addition times large enough)}%
\label{dlarge}
\bea
d \ge d_0 + S_a + r \nn
\eea

\een
where $C'(.,.)$, $C''(.,.)$ are defined in (\ref{defcprime}),
\\ then,
\ben

\item at all $t$, $|\tDelta| \le S_a$, $|\tDelta_e| \le f(S_a+d_0)$ and $|\tT| \le S_0+f(S_a+d_0)$ and the same bounds also hold for $|\Delta|, |\Delta_e|, |T|$ respectively; and

\item for all $t \in [t_j+d_0+S_a-1, t_{j+1}-1]$, $|\tDelta|=0 = |\tDelta_e|$, and thus $\Nhat_t = N_t$ (LS-CS estimate = genie-LS estimate).
\een
\label{stabres}
\end{theorem}

The proof is given in Appendix  \ref{append_lscs_stab_2}. 
{\em Note that in $\min_i a_i$, the $\min$ is taken over $i \in [1,m]$ and same for $\max_i a_i$.}
We now give a simple corollary of Theorem \ref{stabres} (proved in Appendix \ref{corproof}).
\begin{corollary}
If the conditions given in Theorem \ref{stabres} hold,
\ben
\item at all $t$, the LS-CS error satisfies 
\bea
\|(x_t - \xhat_{t})_{\tDelta}\|^2  \sle S_a  \min(M, (d_0+S_a)\max_i a_i)^2 \nn \\
\|(x_t - \xhat_{t})_{\tT}\|^2 \sle   8 {\theta}^2 S_a  \min(M, (d_0+S_a) \max_i a_i)^2  + \nn \\ &&(4{n\lambda^2}/{\|A\|_1^2}) \nn
\label{lscs_err_max}
\eea
with $\theta$ computed at $S_T=S_0+f(d_0+S_a)$, $S_\Delta=S_a$

\item at all $t$, the CS-residual error, $\|x_t - \xhat_{t,\CSres}\|^2$, is bounded by 
\bea
\max(B_0,  C' +  {\theta}^2 C'' S_a  \min(M,(d_0+S_a) \max_i a_i)^2  )  \nn
\label{csres_err_max}
\eea
with $\theta, C', C''$ computed at $S_T=S_0+f(d_0+S_a)$, $S_\Delta=S_a$, and $B_0$ defined in (\ref{defB0}).

\een
\label{stabres_cor}
\end{corollary}

\begin{remark}
Note that the initialization assumption is not restrictive. Denote the bound given by Theorem 1.1 of \cite{dantzig} for $S=S_0-S_a$ by $B1$. It is easy to see that this assumption will hold if the addition threshold at $t=0$ is $\alpha_{init}=\sqrt{B1}$ (ensures no false detects) and if $M > \alpha_{init} + \sqrt{B1}=2\sqrt{B1}$ (ensures all true adds detected). If the noise is small enough, by choosing $n_0$ large enough, we can make $B1$ small enough.%
\\
Even if this cannot be done, our result will only change slightly. The misses can be combined with the new additions at $t=1$. Extras will at most increase the bound on $|T|$ by $f$.
\label{initremark}
\end{remark}

\begin{remark}
By using Corollary \ref{cor2}, instead of Corollary \ref{cor1}, as the starting point for proving the above result, it should be possible to weaken conditions \ref{thetabnd} and \ref{add_incr}. We have not done this here, in order to convey the basic idea in a simpler fashion.
\end{remark}



\begin{table*}[ht!]
\caption{Comparing normalized MSE of CS-residual (with $\lambda=4\sigma$) with that of CS (Dantzig Selector (DS)) with three different $\lambda$'s. We used $m=200$, $|N|=20$, $|\Delta|=|\Delta_e|=2$. Comparison shown for three choices of $n=45,59,100$ in the three tables below.}
\centerline{
\begin{tabular}{|l|l|l|l|l|}
\hline
\hline
($n=45$) & $n$=45 & $n$=45 & $n$=45  & $n$=45 \\
             & $\sigma$=0.04 &  $\sigma$=0.09 &  $\sigma$=0.18 & $\sigma$=0.44 \\
 \hline
DS,$\lambda$=$12\sigma$ & 0.8235   &   0.8952  &    0.9794    &  1.0000 \\
 \hline
DS,$\lambda$=$4\sigma$ & 0.7994   &   0.8320    &  0.8642   &   0.9603 \\
\hline
DS,$\lambda$=$0.4\sigma$  & 0.8071   &   0.8476   &   0.8762  &    1.0917 \\
\hline
CS-residual             &      0.1397   &   0.1685   &   0.2270    &  0.5443 \\
\hline
\hline
\end{tabular}
\hspace{0.05in}
\begin{tabular}{|l|l|l|l|l|}
\hline \hline
($n=59$) & $n$=59 & $n$=59 & $n$=59  & $n$=59 \\
                & $\sigma$=0.04 &  $\sigma$=0.09 &  $\sigma$=0.18 & $\sigma$=0.44 \\
 \hline
DS,$\lambda$=$12\sigma$ & 0.7572   &   0.8402   &   0.9937   &   1.0000 \\
 \hline
DS,$\lambda$=$4\sigma$ & 0.6545  &    0.6759  &    0.7991    &  0.9607  \\
\hline
DS,$\lambda$=$0.4\sigma$ & 0.5375   &   0.5479   &   0.7086    &  1.0525 \\
\hline
CS-residual & 0.0866       &  0.1069  &    0.1800   &   0.4102 \\ 
\hline
\hline
\end{tabular}
\hspace{0.05in}
\begin{tabular}{|l|l|l|}
\hline \hline
($n=100$) & $n$=100 & $n$=100 \\ 
    & $\sigma$=0.04 &  $\sigma$=0.09 \\ 
 \hline
DS,$\lambda$=$12\sigma$ &  0.5856   &   0.8547  \\  
\hline
DS,$\lambda$=$4\sigma$ & 0.2622   &  0.4975   \\ 
\hline
DS,$\lambda$=$0.4\sigma$ & 0.0209  &   0.0929   \\ 
\hline
CS-residual             & 0.0402   &  0.0687 \\  
\hline
\hline
\end{tabular}
}
\label{static_table}
\end{table*}

\Subsection{Discussion and Extensions} 
\label{implications}
Notice that Signal Model \ref{sigmod1} results in {\em bounded SNR and roughly constant signal support size} at all times. Theorem \ref{stabres} and Corollary \ref{stabres_cor} show that under Signal Model \ref{sigmod1} and under the initialization assumption (made only for simplicity), if
\ben
\item the noise is bounded and $n$ is large enough so that condition \ref{measmodass} holds,
\item the addition/deletion thresholds are appropriately set (condition \ref{algo_thresh}),
\item for a given noise bound and $n$, the smallest rate of coefficient magnitude increase is large enough (condition \ref{additioncond}) and the smallest constant coefficient magnitude is also large enough (conditions \ref{nofd_const}, \ref{nofd_decr}),
\item and the delay between addition times is larger than the ``detection delay" (which in turn depends on the magnitude increase rate), i.e. condition \ref{dlarge} holds,
\een
then,
\ben
\item the number of misses, $|\tDelta| \le S_a$, and the number of extras, $|\tDelta_e| \le f(S_a+d_0)$ and the same bounds hold for $|\Delta|$, $|\Delta_e|$ (here $d_0 \le d$ is the smallest integer for which the conditions of Theorem \ref{stabres} hold), i.e. ``stability" holds; 

\item within a finite ``detection delay", $d_0+S_a-1$, all new additions get detected and not falsely deleted ($|\tDelta|=0$), and the extras get deleted ($|\tDelta_e|=0$), i.e. the LS-CS estimate becomes equal to the genie-LS estimate;
\item  and the LS-CS error and the CS-residual error are bounded by time-invariant values.
\een

From Assumption \ref{initass} (given in Sec. \ref{probdef}), $S_a  \ll  S_0$. 
When $n$ is large enough (as required above), it is easy to set $\alpha$ so that $f$ is small, e.g. in our simulations the average $f$ was often less than 1 while $S_0=20$. With a fast enough signal increase (as required above), $d_0$ will also be small. Thus we can claim that $|\tDelta|$ and $|\tDelta_e|$ will be bounded by a small value compared to the signal support size, $S_0$, i.e. ``stability" is meaningful.

Under the above assumptions, compare our requirements on $n$ (condition \ref{measmodass}) to those of the CS error bound \cite{dantzig}, which needs $S_0 \le \sinf$. 
The comparison is easier to make if we slightly modify the definition of $\sinf$ to be the largest $S$ for which $\delta_{2S} < 1/2$ and $\delta_{2S}+\theta_{S,2S}<1$ (this will imply that $2\sinf \le \sone$). Clearly $S_a \le \sinf$ is much weaker than $S_0 \le \sinf$. Also, $S_0 + f(d_0+S_a) \le \sone$ is weaker than $2S_0 \le \sone$. Finally, if $f,d_0,S_a$ are small enough, condition \ref{thetabnd} is also weaker.%


Notice that our signal model assumes that support changes occur every $d$ time instants. This may be slightly restrictive. But it is necessary in order to answer our second question (do the support errors ever become zero). If we do not care about answering this, we can assume a signal model with $d=1$ and modify our arguments to still ensure stability. But the support errors may never become zero. We do this in \cite{icassp10}. 


Also, note that if $r$ is large (slow rate of decrease),  condition \ref{nofd_decr} becomes difficult to satisfy. If we remove this, we may not be able to prevent false deletion of the decreasing coefficients when they become too small (go below $\alpha_{del}+\frac{2\sqrt{n}\lambda}{\|A\|_1}$). But since they are small, this will increase  the CS-residual error at the next time instant only slightly. With small changes to our arguments, it should be possible to still prove stability.

\Section{Numerical Experiments}
\label{sims}

In Sec. \ref{staticsim}, we study a static problem and compare CS-residual error with that of CS. In Sec. \ref{stabsims}, we
verify the stability result of Theorem \ref{stabres}. In Sec. \ref{lowsnrsims}, we simulate lower SNRs and faster additions. In all these simulations, $A$ was random-Gaussian. We averaged over 100 simulations (noise and signal supports for all times randomly generated) for all the time-series         simulations and over 50 for the static one. In Sec. \ref{dynmriapp}, we show a dynamic MRI reconstruction example.
All our code used \texttt{CVX}, \url{www.stanford.edu/~boyd/cvx/}.



\Subsection{Comparing CS-residual with CS}
\label{staticsim}
We simulated a single time instant reconstruction problem (reconstruct $x$ from $y:=Ax+w$) with $m=200$,  $|N|=20$, and with $|\Delta| = 0.1|N|=2=|\Delta_e|$. The noise $w$ was zero mean i.i.d Gaussian. The nonzero signal values, $x_N$, were i.i.d. $\pm 1$ with equal probability.
The sets $N$, $\Delta \subseteq N$ and $\Delta_e  \subseteq N^c$ were uniformly randomly generated each time.
We used four different noise standard deviations ($\sigma = 0.0439*[1,2,4,10]$) and three different choices of $n$ (45, 59, 100).  In Table \ref{static_table}, we compare the normalized MSE (NMSE) of CS-residual output with that of CS. CS (Dantzig selector) was run with different choices of $\lambda$ while for CS-residual we fixed $\lambda = 4 \sigma$. Except when $n=100$, in all other cases CS-residual outperforms CS significantly. For $n=100$ (large $n$), if $\sigma = 0.04$, CS (with smallest $\lambda$) is better, and if $\sigma = 0.09$, both are similar.


A few other observations. (1) When $n$ is small, the best CS error occurs when we run it with the smallest $\lambda$. Smaller $\lambda$ reduces the size of the feasible set and thus the $\ell_1$ norm of the minimizer, $\xhat$, is larger, i.e. more of its elements are nonzero (if $\lambda$ is too large, $\xhat=0$ will be feasible and will be the solution). (2) We also compared the CS-residual error with the error of the final LS-CS output (not shown). Only when CS-residual error was small, the support estimation was accurate and in this situation the final LS-CS error was much smaller.



\begin{figure}
\vspace{-0.1in}
\centerline{
\epsfig{file = 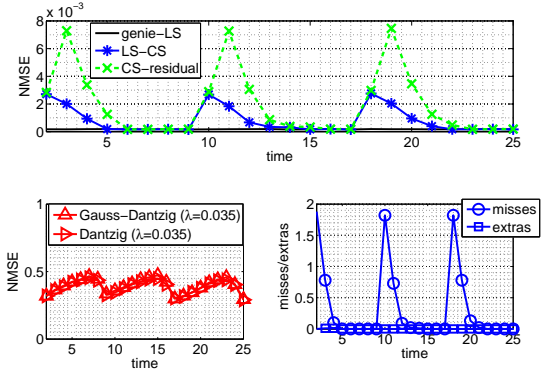, width=8cm}
}
\vspace{-0.15in}
\caption{\small{{Verifying the stability result.} 
}}
\vspace{-0.1in}
\label{stability_fig}
\end{figure}

\Subsection{Verifying LS-CS stability}
\label{stabsims}
In Fig. \ref{stability_fig}, we verify the stability result. We simulated Signal Model \ref{sigmod1} with $m=200$, $S_0=20$, $S_a=2$ and with $d=8$, $r=2$, $M=3$. Half the $a_i$'s were 0.5, the other half were 0.25. We used $n=59$ and noise was $unif(-c,c)$ with $c=0.0528$. The LS-CS algorithm used  $\lambda=c\|A\|_1=0.35$, $\alpha=c$ and $\alpha_{del}=2.28c$. We assumed that the initialization condition holds, i.e. we started LS-CS with $\Nhat_0 = N_0$.
In all 100 simulations, the number of misses and extras became exactly zero within $d_0+S_a-1=4$ time units of the addition time, i.e.  the LS-CS estimate became equal to that of the genie-LS. Thus $d_0$ was at most 3 in the simulations. The NMSE of LS-CS is $\le$ 0.4\% while that of CS with small $\lambda$, is 30-40\%.



\begin{figure}
\centerline{
\subfigure[Low SNR, Slow adds, $n$=59]{
\label{kfcs_better_59}
\epsfig{file = 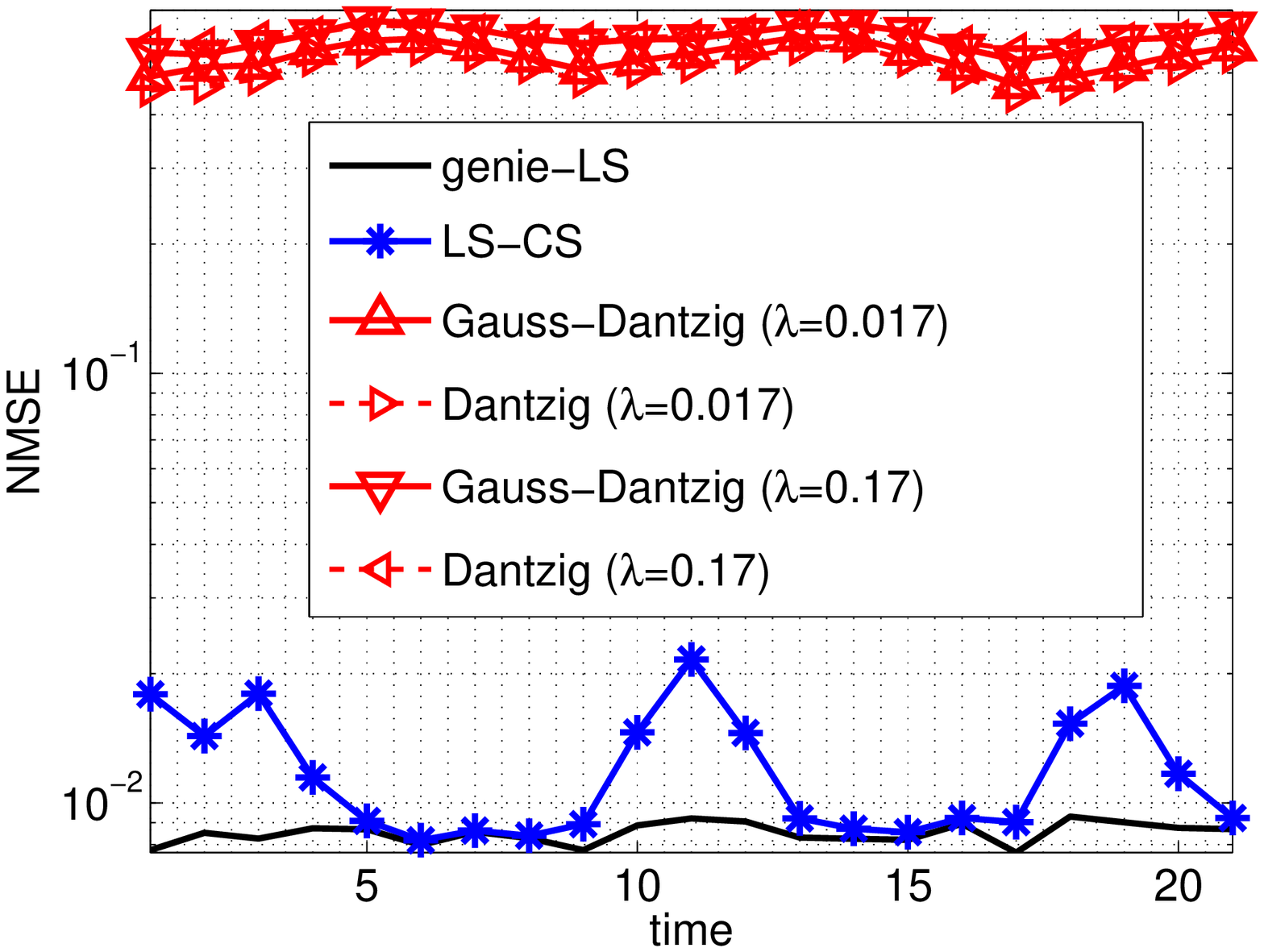, height = 3cm,width=4cm} 
}
\subfigure[Low SNR, Slow adds, $n$=59]{
\label{kfcs_better_59_support}
\epsfig{file = 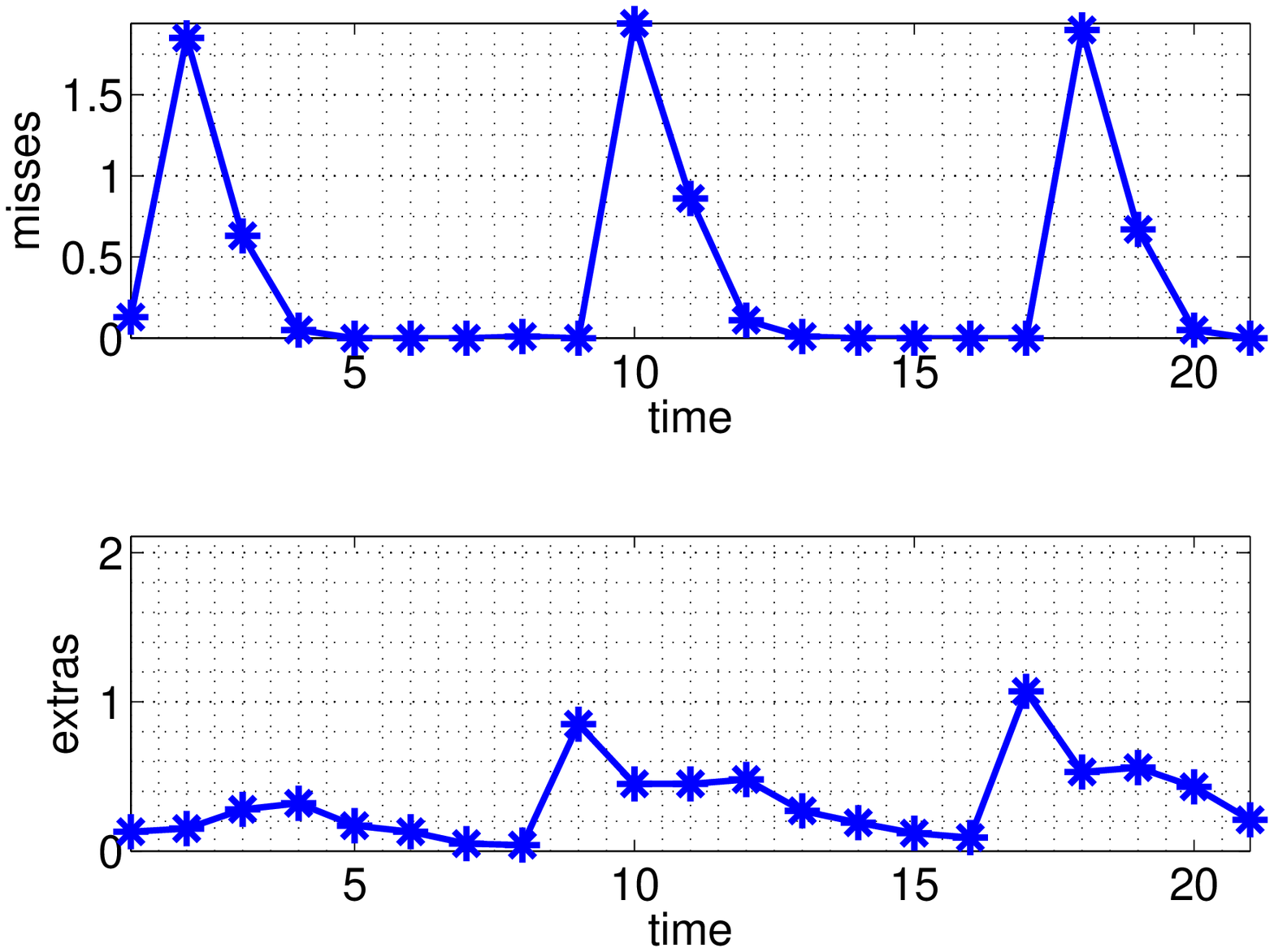, height = 3cm,width=4cm}
}
}
\centerline{
\subfigure[Low SNR, Fast adds, $n$=59]{
\label{fast_adds_dels}
\epsfig{file = 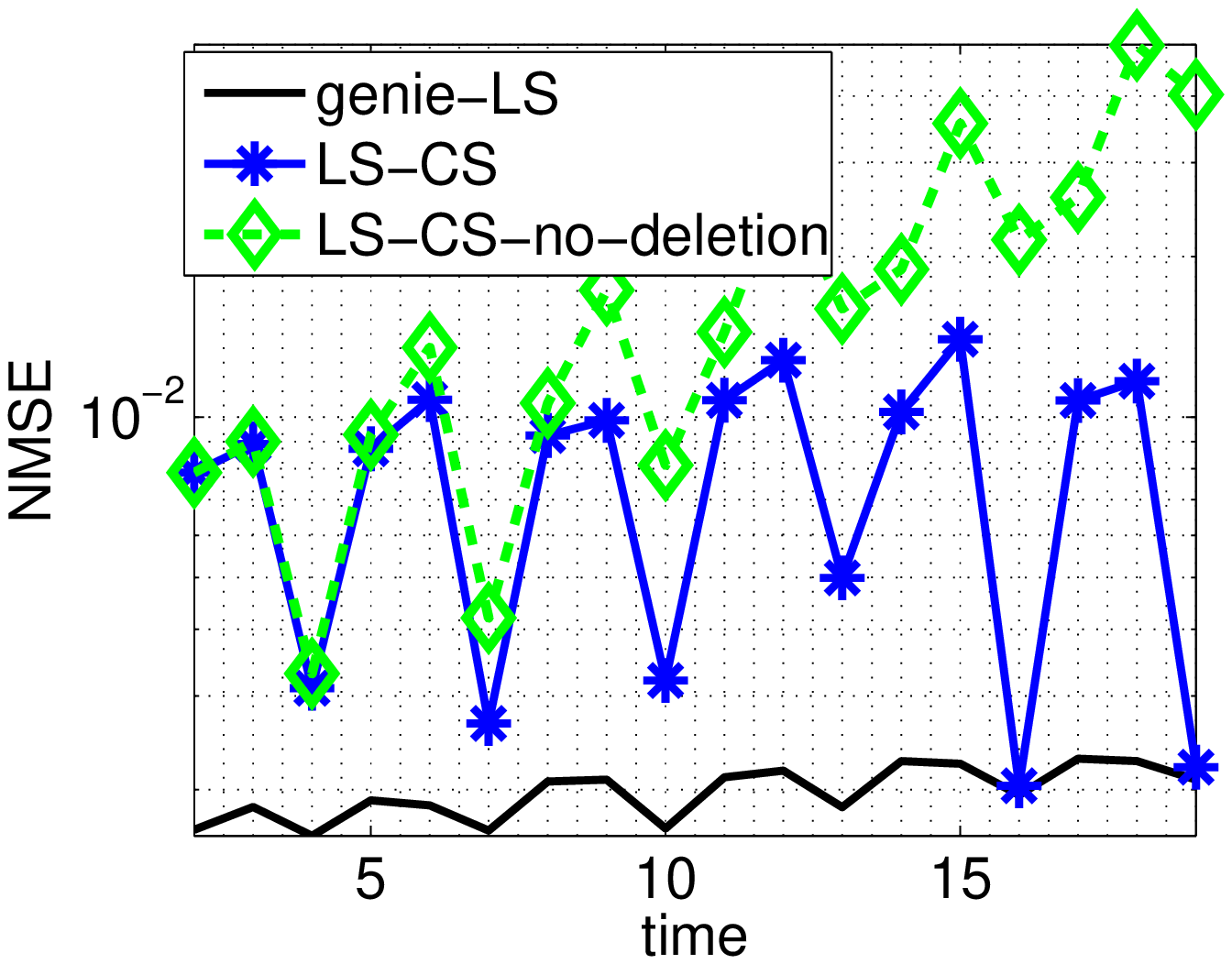, height = 3cm,width=4cm}
}
\subfigure[Low SNR, Fast adds, $n$=59]{
\label{fast_adds_dels_support}
\epsfig{file = 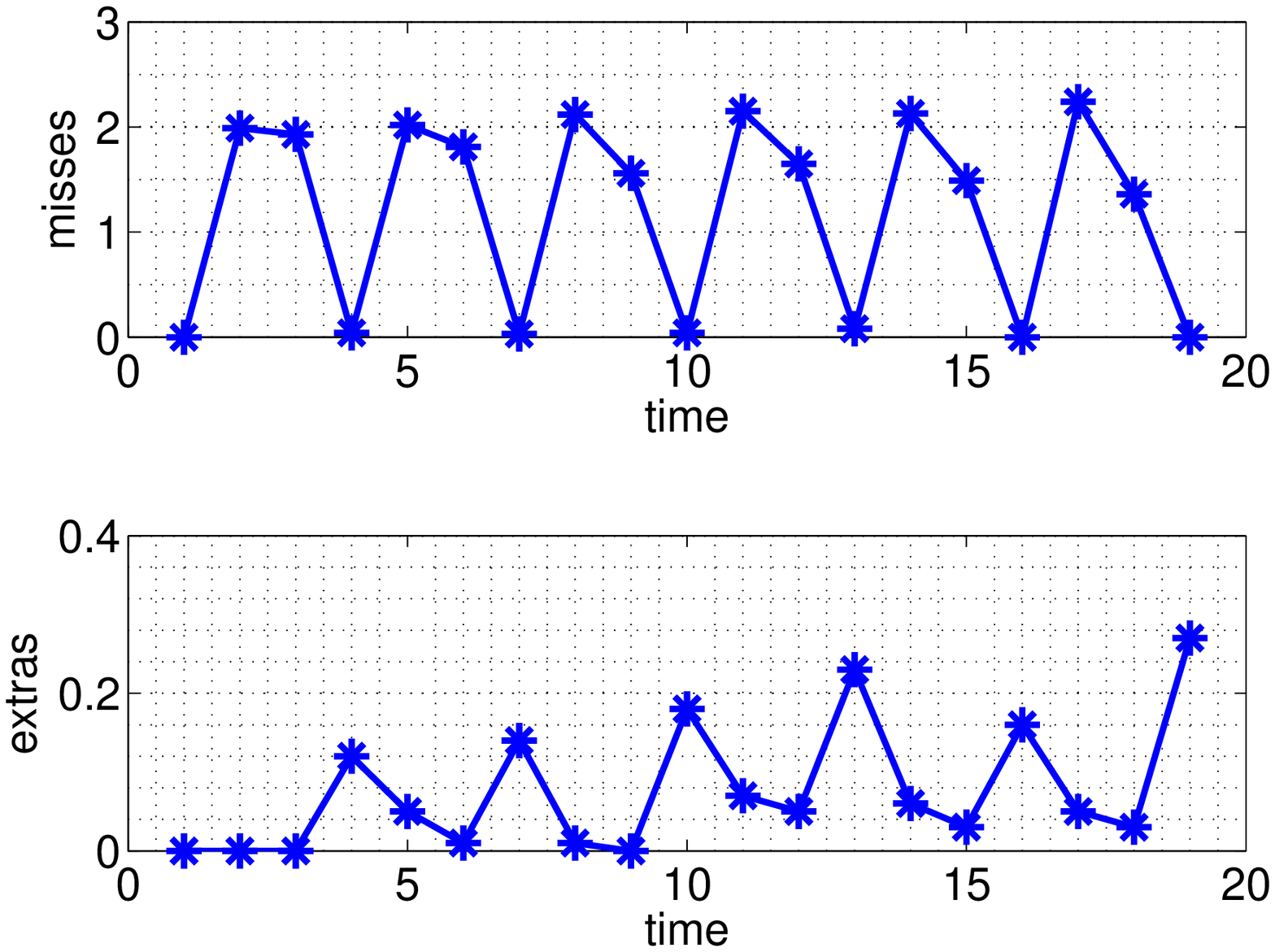, height = 3cm,width=4cm}
}
}
\vspace{-0.1in}
\caption{\small{Lower SNRs and Faster additions. LS-CS-no-deletion refers to LS-CS without deletion step. y axis is log scale in Fig. \ref{kfcs_better_59},\ref{fast_adds_dels}.%
}}
\vspace{-0.1in}
\label{lscs_kfcs}
\end{figure}

\Subsection{Lower SNR and faster additions}
\label{lowsnrsims}
Next we ran two sets of simulations with much lower SNRs - slow-adds and fast-adds. Slow-adds used $d=8$, while fast-adds had faster additions, $d=3$. In all simulations, $m=200$, $S_0=20$, $S_a=2$ and the noise is $unif(-c,c)$. Also, we used a smaller $\lambda$,  $\lambda = c\|A\|_1/2$ since it encourages more additions.

We define two quantities: minimum average signal to noise ratio (min-SNR) and  maximum average signal to noise ratio (max-SNR). Min (max) SNR is the ratio of minimum (maximum) average signal magnitude to the noise standard deviation. For $unif(-c,c)$ noise, the standard deviation is $c/\sqrt{3}$. Min-SNR, which occurs right after a new addition, decides how quickly new additions start getting detected (decides $d_0$). Max-SNR decides whether $|\Delta|$ becomes zero before the next addition. Both also depend on $n$ of course.

For the previous subsection (Fig. \ref{stability_fig}), $c/\sqrt{3} = 0.03$.  Minimum average signal magnitude was $(0.5+0.25)/2=0.375$ while maximum was $(M+d\min_i a_i)/2 = (3+8*0.25)/2 = 2.5$. {\em Thus min-SNR was 12.3 while max-SNR was 82.}

In slow-adds (Fig. \ref{kfcs_better_59},  \ref{kfcs_better_59_support}), we use $n=59$, $c=0.1266$ and Signal Model \ref{sigmod1} with $a_i = 0.2$, $M=1$, $d=8$ and $r=3$. Thus {\em min-SNR was $0.2*\sqrt{3}/0.1266 = 2.73$ while max-SNR was $1*\sqrt{3}/0.1266=13.7$ (both are much smaller than 12.3 and 82 respectively)}.
LS-CS used $\lambda = 0.176$, $\alpha=c/2 =0.06= \alpha_{del}$. Also, it restricted maximum number of additions at a time to $S_a+1$. We also evaluated our assumption that CS at $t=0$ done with large enough $n_0$ finds the support without any error. With $n_0=150$, this was true 90\% of the times, while in other cases there were 1-2 errors. 
 Notice the following.
(1) Most additions get detected within 2 time units and there are occasionally a few extra additions. This is because we set $\alpha=\alpha_{del}=c/2$ (both very low).
(2) As long as $A_T'A_T$ remains well-conditioned, a few extras do not increase the error visibly above that of the genie-LS. Notice from the plots that even when LS-CS $\approx$ genie-LS, the average extras, $|\tDelta_e|$, are not zero. (3) LS-CS error (NMSE) is stable at 2.5\% while the CS errors are much larger at 40-60\%.



In fast-adds (Fig. \ref{fast_adds_dels}, \ref{fast_adds_dels_support}), we use $n=59$, $c=0.0528$ and a slightly modified Signal Model \ref{sigmod1} with $a_i = 0.2$, $M=1$, $d=3$ and $r=2$. Thus {\em min SNR was $0.2*\sqrt{3}/0.0528 = 6.6$ while max SNR was $0.6*\sqrt{3}/0.0528= 19.7$.} Both are smaller than the stability simulation, but larger than the slow-adds simulations. This was needed because in this case the delay between addition times was only 3, and so quick detection was needed to ensure error stability. 
LS-CS used $\lambda = 0.176$, $\alpha=c = 0.05 = \alpha_{del}$ and maximum additions per unit time of $S_a=2$. 
LS-CS error (NMSE) is still stable at 1\%.

\newcommand{\kron}{\otimes}

\subsection{Dynamic MRI reconstruction example}
\label{dynmriapp}

To address a reviewer comment, in Fig. \ref{dynmri}, we show the applicability of LS-CS to accurately reconstruct a sparsified cardiac image sequence from only 35\% (simulated) MRI measurements. Detailed comparisons for actual (not sparsified) image sequences, using practical MR data acquisition schemes, and with using BPDN are given in \cite{chenlu_tip}.

For Fig. \ref{dynmri}, the sparsity basis was the two-level Daubechies-4 2D DWT. Images were 32x32 ($m=1024$) and were sparsified by retaining the largest magnitude DWT coefficients that make up 99.5\% of the total image energy and computing the inverse DWT. The support size of the sparsified DWT vector varied between 106-110, and the number of additions to (or removals from) the support from any $t-1$ to $t$ varied between 1-3.
Denote the 1D DWT matrix by $W$ and the DFT matrix by $F$. Then $\Phi = W \kron W$ and the measurement matrix, $H = M_{rs}(F \kron F)/32$ where $M_{rs}$ is an $n \times m$ random row selection matrix and $\kron$ denotes the Kronecker product. We used $n=0.35m$ and $n_0=0.8m$. Noise was zero mean i.i.d. Gaussian with variance $\sigma^2 = 0.125$. Both LS-CS and CS used $\lambda=1.5 \sigma$. We also tried running CS with smaller values of $\lambda$:  $\lambda=0.15 \sigma$ and $\lambda=0.3 \sigma$, but these resulted in (\ref{simplecs}) being infeasible.


\begin{figure}
\centerline{
\subfigure[\small{NMSE comparison (y axis is log scale)}]{
\begin{tabular}{c}
\includegraphics[height=3cm,width=7cm]{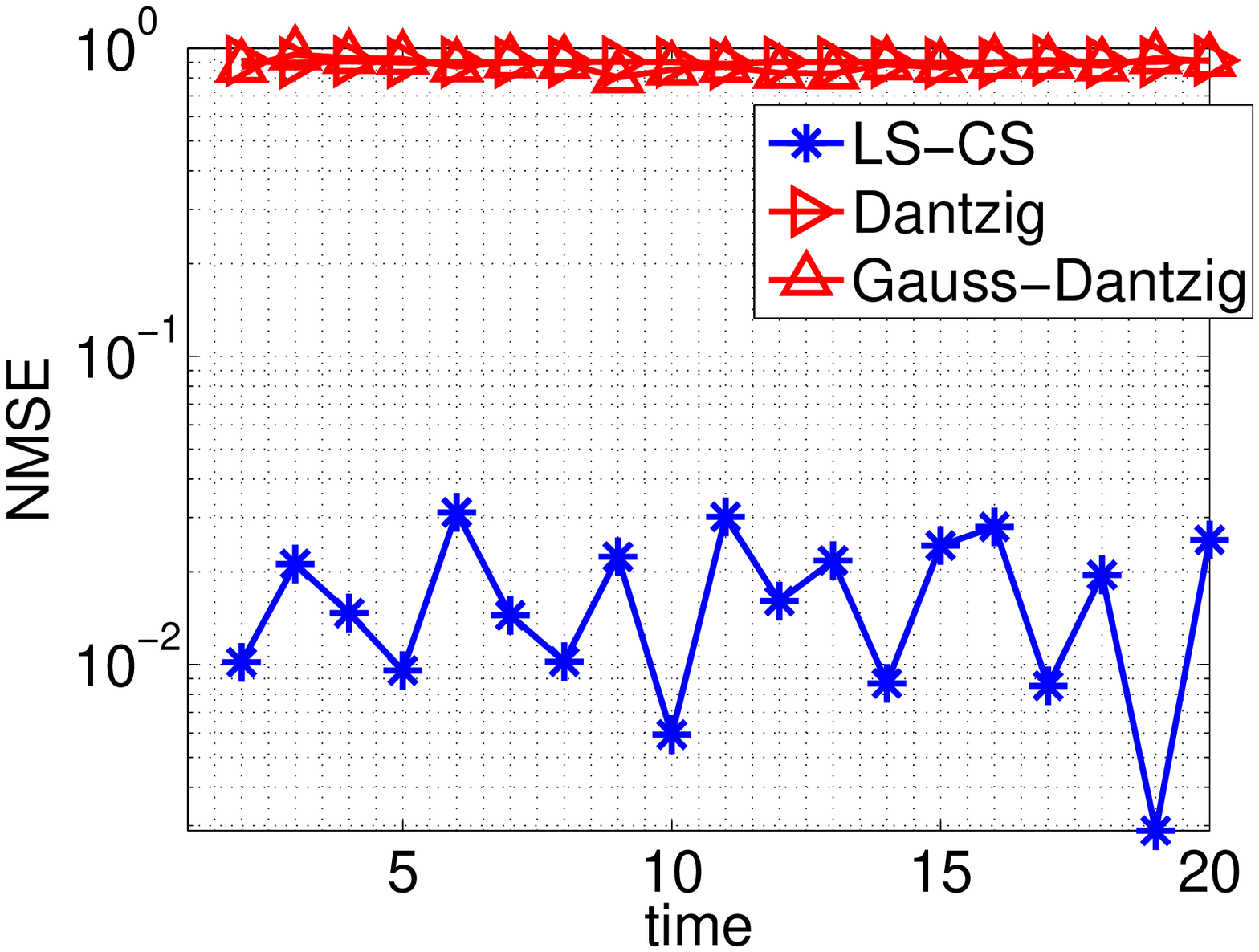}
\end{tabular}
}
}
\centerline{
\subfigure[\small{Frames 2, 11, 20: original and reconstructed}]{
\begin{tabular}{cc}
\includegraphics[height=1cm]{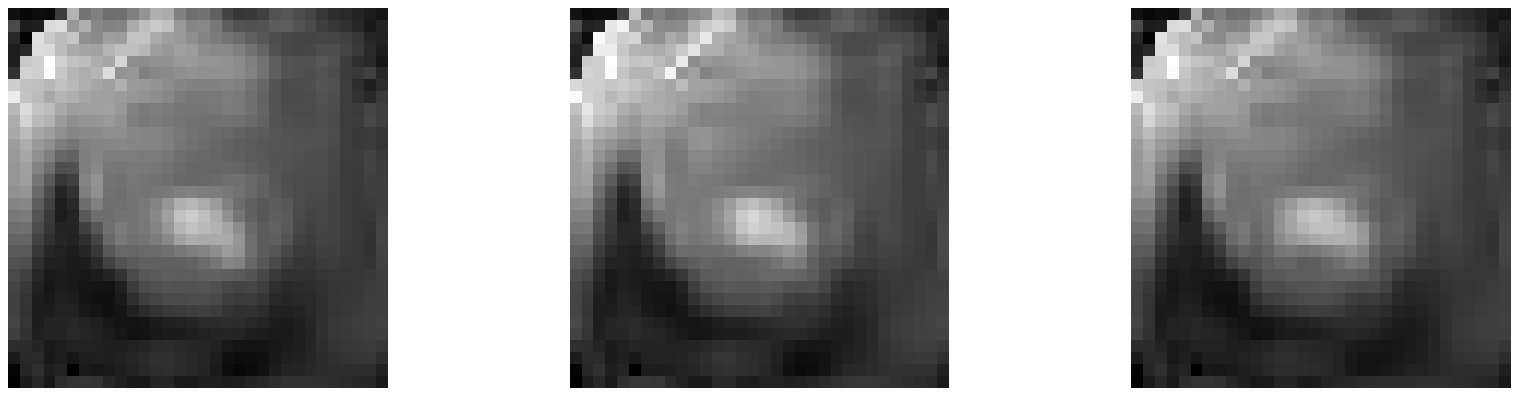} &  
\includegraphics[height=1cm]{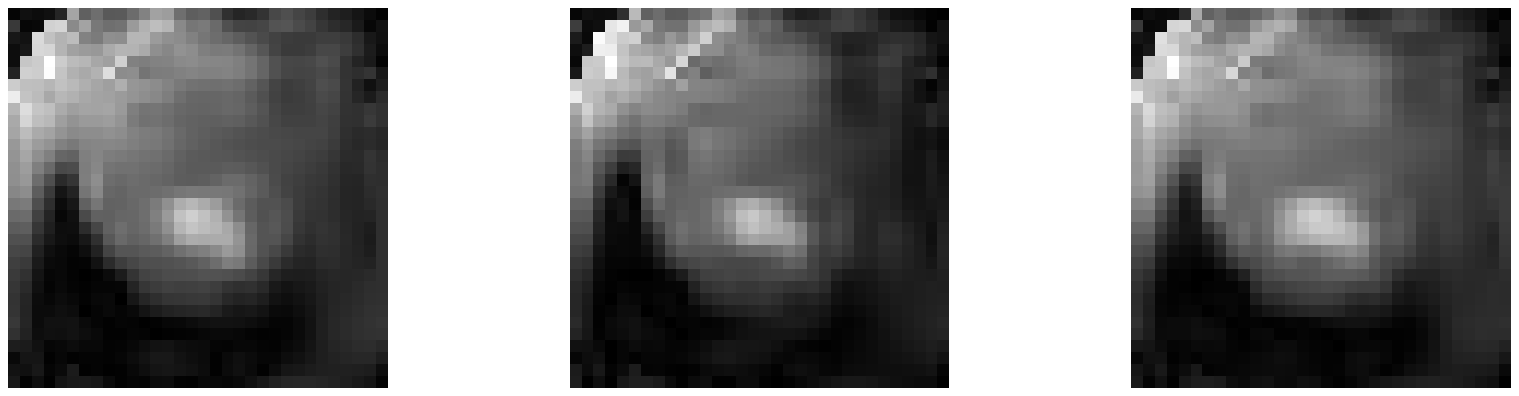} \\
\small{Original} & \small {LS-CS reconstruction} \\
\includegraphics[height=1cm]{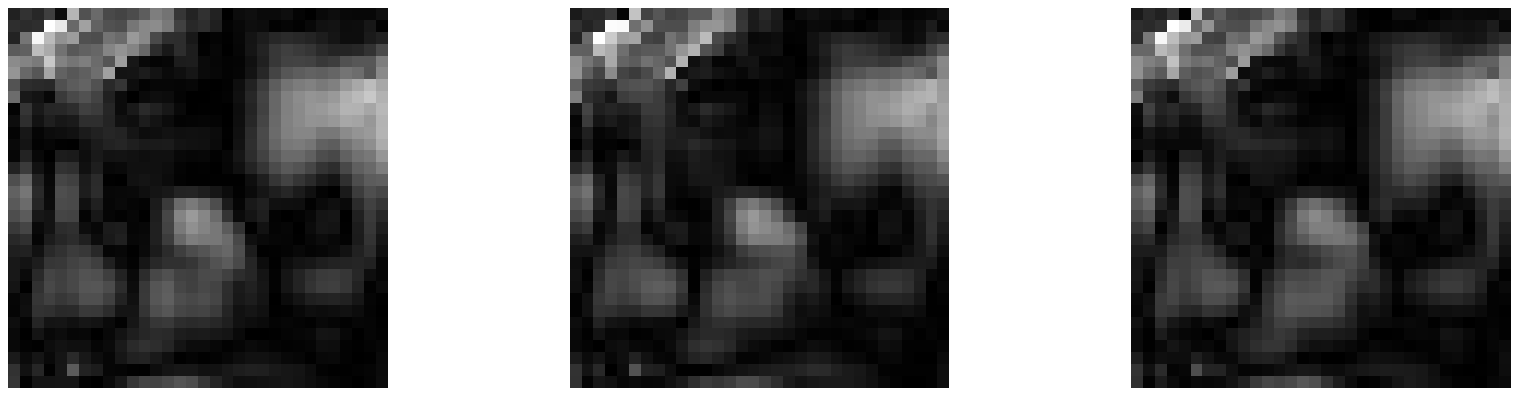} &
\includegraphics[height=1cm]{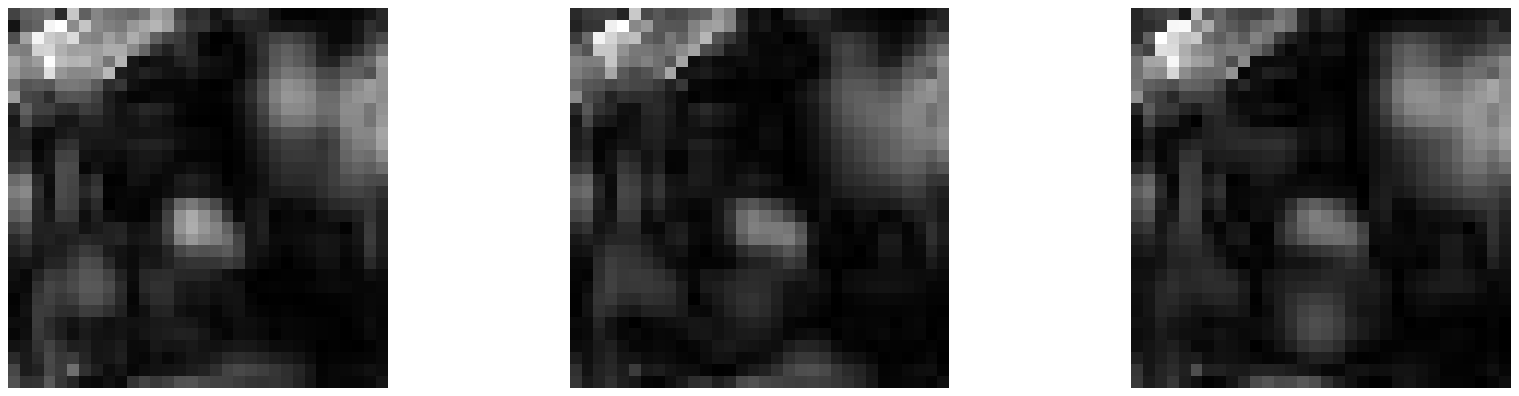} \\
\small{CS reconstruction} & \small {Gauss-CS reconstruction} \\
\end{tabular}
}
}
\vspace{-0.1in}
\label{dynmri}
\caption{\small{Dynamic MRI. Reconstructing a sparsified cardiac sequence.
}}
\vspace{-0.2in}
\label{dynmri}
\end{figure}


 \vspace{-0.1in}
\Section{Conclusions}
\label{conclusions}
 \vspace{-0.05in}

We formulated the problem of recursive reconstruction of sparse signal sequences from noisy observations as one of noisy CS with partly known support (the support estimate from the previous time serves as the ``known" part). Our proposed solution, LS CS-residual (LS-CS), replaces CS on the raw observation by CS on the LS residual, computed using the known part of the support. 
We obtained bounds on CS-residual error. 
When the number of available measurements, $n$, is small, we showed that our bound is much smaller than the CS error bound if $|\Delta|, |\Delta_e|$ are small enough. We used this bound to show the stability of LS-CS over time. By ``stability" we mean that $|\Delta|, |\Delta_e|$ remain bounded by time-invariant values. Extensive simulation results backing our claims are shown.



An open question is how to prove stability for a stochastic signal model that uses a random walk model with drift given by the current model for coefficient increase/decrease while using a (statistically) stationary model for ``constant" coefficients, and that assumes a prior on support change, e.g. a modification of the model of \cite{schniter}. 
Finally, in this work, we did not study exact reconstruction using much fewer noise-free measurements. We do this in  \cite{isitmodcs}.

\appendix

\subsection{CS-residual Bound: Proof of Lemma \ref{thm13} and Theorem \ref{thm1}}
\label{thm13_proof}

\subsubsection{Proof of Lemma \ref{thm13}}
The proof is a modification of the proof of Theorem 1.3 given in \cite{dantzig}. Let $\delta \equiv \delta_{2S}$, $\theta \equiv \theta_{S,2S}$.
Let $\zetahat = \zeta + h$. 
Let $T_0 \subseteq T_{nz}$ be a size $S$ subset with $S \le \min(\sinf,S_{nz})$ and let $\rest = T_{nz} \setminus T_0$.
$\|w\|_\infty \le \frac{\lambda}{\|A\|_1}$ implies that $|{A_i}'w| \le \|A_i\|_1 \|w\|_\infty \le \lambda$. Thus eq. (3.1) of \cite{dantzig} holds with probability (w.p.) 1 and so $\zeta$ is feasible. Thus,
\bea
\label{hT0c}
\|(h)_{T_0^c}\|_1 \sle \|(h)_{T_0}\|_1 + 2 \|(\zeta)_{T_0^c}\|_1  \\  
\label{eq33}
\|A'A h \|_\infty \sle 2 \lambda
\eea
The second equation is eq (3.3) of \cite{dantzig}. The first follows by simplifying $\|\zetahat\|_1 \le \|\zeta\|_1$ \cite{dantzig}.

Recall that $\sinf$ is the largest value of $S$ for which $\delta+\theta < 1$. Thus we can apply Lemma 3.1 of \cite{dantzig} for any $S \le \sinf$.
Let $T_1$ contain the indices of the $S$ largest magnitude elements of $h : = \zetahat - \zeta$ outside of $T_0$. Let $T_{01} := T_0 \cup T_1$. Thus $|T_{01}| = 2S$ and $\|h_{T_0}\|_k \le \|h_{T_{01}}\|_k$ for any $\ell_k$ norm.
Apply Lemma 3.1 of \cite{dantzig} and use (\ref{hT0c}) and (\ref{eq33}) to upper bound its first inequality. Then use $\|h_{T_0}\|_1/\sqrt{S} \le  \|h_{T_0}\| \le \|h_{T_{01}}\|$ to simplify the resulting inequality, and then use $(a+b)^2 \le 2a^2 + 2b^2$ to square it. Finally, use $\|(\zeta)_{T_0^c}\|_1=\|(\zeta)_{\rest}\|_1$ to get
\bea
\|h_{T_{01}}\|^2 \sle \frac{16 S \lambda^2}{(1-\delta - \theta)^2} + \frac{8 \theta^2 \| \zeta_{\rest}\|_1^2}{(1-\delta - \theta)^2 S} \ \
\label{hT01}
\eea
Using $(a+b)^2 \le 2a^2 + 2b^2$ to simplify the square of (\ref{hT0c}); using the resulting bound in the second inequality of Lemma 3.1 of \cite{dantzig}; and then finally using (\ref{hT01}), we get
\bea
\|h\|^2 \sle \frac{48 S \lambda^2 }{(1-\delta - \theta)^2} + (8 + 24 \frac{\theta^2}{(1-\delta - \theta)^2})\frac{\|\zeta_{\rest}\|_1^2}{S} \ \ \ \ \ \ \ \
\eea
Since $\rest = T_{nz} \setminus T_0$ and $|T_0|=S$, thus $|\rest| = S_{nz} - S$. Thus $\|\zeta_{\rest}\|_1^2 \le (S_{nz} - S)\|\zeta_{\rest}\|^2$. This gives our result which holds for any set $T_0 \subseteq T_{nz}$ of size $S \le \min(\sinf,S_{nz})$. $\blacksquare$

\subsubsection{Proof of Theorem \ref{thm1}}
\label{thm1_proof}
The  result follows by applying Lemma \ref{thm13} with $\zeta = \beta$, $S_{nz} = |T|+|\Delta|$ and picking the set $\rest$ of size $|T|+|\Delta|-S$ as follows. For $S \ge |\Delta|$,  pick $\rest \subseteq T$ of size $|T|+|\Delta|-S$ and bound $\|\beta_\rest\|$ by $\|\beta_T\|$. Use (\ref{betabound}) to bound $\|\beta_T\|$, and use $\delta_{|T|} < 1/2$ to simplify the final expression.
For $S < |\Delta|$, pick the set $\rest$ as the set $T$ union with $|\Delta|-S$ smallest elements of $x_\Delta$.  Finally use
 $\xhat_{\CSres} = \betahat + \xhat_{\text{init}}$ and $\beta = x - \xhat_{\text{init}}$ to get $\beta - \betahat = x - \xhat_{\CSres}$. Lastly, from the definitions, $|T|=|N|+|\Delta_e|-|\Delta|$.

\Subsection{LS-CS Stability: Proofs of the Key Lemmas for Theorem \ref{stabres}}
\label{append_lscs_stab_1}
The proofs of the three lemmas essentially follow from Corollary \ref{cor1} and the following simple facts.
\ben
\item An $i \in \Delta$ (an undetected element) will definitely get detected at current time if $x_i^2 > 2\alpha^2 + 2\|x - \xhat_\CSres\|^2$~  \footnote{An $i \in \Delta$ will get detected if $|(\xhat_{\CSres})_i |  > \alpha$. Since $|(\xhat_{\CSres})_i| \ge |x_i| - |x_i - (\xhat_{\CSres})_i| \ge  |x_i| - \|x - \xhat_\CSres\|$, this holds if $|x_i| > \alpha + \|x - \xhat_\CSres\|$. This, in turn, holds if $x_i^2 > 2 \alpha^2 + 2\|x - \xhat_\CSres\|^2$.}.%
\label{det1}

\item An $i \in (\tT_\dett \setminus \tDelta_{e,\dett})$ (a nonzero element of the current detected set) will definitely not get falsely deleted at the current time if $x_i^2 > 2 \alpha_{del}^2 + 2\|(x - \xhat_\dett)_{\tT_\dett}\|^2$.
\label{nofalsedel1}

\item All $i \in \tDelta_{e,\dett}$ (a zero element of the current detected set) will get deleted if $\alpha_{del}^2 \ge \|(x - \xhat_\dett)_{\tT_\dett}\|^2$.
\label{truedel1}

\item If $\|w\|_\infty \le \lambda/\|A\|_1$ and $|\tT_\dett| \le \sone$, then
\label{errls1}
$
\|(x - \xhat_\dett)_{\tT_\dett}\|^2 \le (4n\lambda^2/\|A\|_1^2) + 8 {\theta_{|\tT_\dett|,|\tDelta_\dett|}}^2 \|x_{\tDelta_\dett}\|^2 \nn 
\le  (4n\lambda^2/\|A\|_1^2) + 8 {\theta_{|\tT_\dett|,|\tDelta_\dett|}}^2 |\tDelta_\dett| \|x_{\tDelta_\dett}\|_\infty^2
$

\item The bound in fact \ref{errls1} is non-decreasing in $|\tT_\dett|$ and $|\tDelta_\dett|$.%
\label{nondec}
\een

{\em Proof of Lemma \ref{largest}: }  From  Corollary \ref{cor1} and the fact that $\|x_\Delta\|^2 \le |\Delta| (x_\Delta)_{(1)}^2$, if $\|w\|_\infty \le \lambda/\|A\|_1$, $|T| \le \sone$ and $|\Delta| \le \sinf$, then $\| x -\xhat_{\CSres} \|^2 \le C'+ C'' \theta^2 |\Delta| (x_\Delta)_{(1)}^2$ with $C',C'',\theta$ computed at $|T|,|\Delta|$. $C', C''$ are defined in (\ref{defcprime}).

Using fact \ref{det1} from above, the largest undetected element, $(x_\Delta)_{(1)}$, will definitely get detected at the current time if $(x_\Delta)_{(1)}^2  > 2\alpha^2 + 2C'+ 2C'' \theta^2 |\Delta| (x_\Delta)_{(1)}^2$. Clearly this holds if
$2{\theta}^2 {|\Delta| C''}<1$ and $\frac{2\alpha^2 + 2C'}{1- 2 \theta^2 |\Delta| C''} < (x_\Delta)_{(1)}^2$.
If it is only known that $|T| \le S_T$ and $|\Delta| \le S_\Delta$ then our conclusion will hold if the maximum of the left hand sides (LHS) over $|T| \le S_T$ and $|\Delta| \le S_\Delta$ is less than the right side. This gives the lemma. The LHS of the first inequality is non-decreasing in $|T|,|\Delta|$ and hence is maximized for $S_T,S_\Delta$. The LHS of the second one is non-decreasing in $|T|$ but is not monotonic in $|\Delta|$.

%

{\em Proof of Lemma \ref{nofalsedelscond}: } It follows from facts \ref{nofalsedel1}, \ref{errls1} and \ref{nondec}.

{\em Proof of Lemma \ref{truedelscond}: } It follows from facts \ref{truedel1}, \ref{errls1} and \ref{nondec}.

\subsection{LS-CS Stability: Proof of Theorem \ref{stabres}}
\label{append_lscs_stab_2}
 Let $t_0=0$ (call it the zeroth addition time). The first addition time, $t_1 = 1$. We prove Theorem \ref{stabres} by induction. At $t=t_0=0$, all the $S_0-S_a$ coefficients are correctly detected (according to the initialization condition), and thus $|\tDelta|=|\tDelta_e|=0$ and  $|\tT|=|N|=S_0-S_a$. Thus for the initial interval $t \in [t_0, t_1-1]$, our result holds. This proves the base case. 
Now for the induction step, assume that
\begin{ass}[induction step assumption]
The result holds for all $t \in [t_{j-1}, t_j-1]$. Thus at $t=t_j-1$, $|\tDelta|=|\tDelta_e|=0$ and $|\tT|=|N|=S_0-S_a$.
\label{inducass}
\end{ass}
Then prove that the result holds for $t \in [t_j, t_{j+1}-1]$. The following facts will be frequently used in the proof.
\ben

\item Recall that $t_{j+1} = t_j  + d$. Also, coefficient decrease of the elements of $\rem$ begins at $t_{j+1}-r = t_j + d - r$ and the coefficients get removed at $t_{j+1}-1$. Since $d \ge d_0+S_a + r$ (condition \ref{dlarge} of the theorem), thus, coefficient decrease does not begin until $t_j + d_0 + S_a$ or later.
\label{coef_dec_time}


\item At all $t \in [t_j, t_{j+1}-2]$, $|N| = S_0$, while at $t=t_{j+1}-1$, $|N|=S_0-S_a$. Also, there are $S_a$ additions at $t=t_j$ and none in the rest of the interval $[t_j, t_{j+1}-1]$. There are $S_a$ removals at $t=t_{j+1}-1$, and none in the rest of the interval before that.

\item $\Delta_t \subseteq \tDelta_{t-1} \cup (N_t \setminus N_{t-1})$ and $\Delta_{e,t} \subseteq \tDelta_{e,t-1} \cup (N_{t-1} \setminus N_t)$.
If there are no new additions, $\Delta_t = \tDelta_{t-1}$. Similarly, if there are no new removals, $\Delta_{e,t} = \tDelta_{e,t-1}$. 

\een

The induction step proof follows by combining the results of the following six claims. In each claim, we bound $|\tDelta|$, $|\tDelta_e|$, $|\tT|$ in one of the sub-intervals shown in Fig. \ref{stabfig}. Using the last two facts above, the bounds for $|\Delta|$, $|\Delta_e|$, $|T|$ follow directly.

\begin{claim}
At all $t=t_j+i$, for all $i=0,1,\dots d_0-1$, $|\tDelta| \le S_a$, $|\tDelta_e| \le (i+1)f$,  $|\tT| \le S_0+ (i+1)f$.
\end{claim}

{\em Proof: } We prove this by induction. Consider the base case, $t=t_j$. At this time there are $S_a$ new additions and $|N|=S_0$. Using Assumption \ref{inducass} (induction step assumption), $|\Delta|=S_a$, $|\Delta_e|=0$. In the detection step, 
$|\tDelta_\dett| \le |\Delta|= S_a$ and so $\|x_{\tDelta_\dett}\|_\infty \le \min(M, (a_\add)_{(1)})$.
There are at most $f$ false detects (condition \ref{algo_thresh}), so that $|\tDelta_{e,\dett}| \le 0+ f$.  Thus, $|\tT_\dett| = |N| + |\tDelta_{e,\dett}| - |\tDelta_\dett| \le S_0 + f$.

The smallest constant coefficient has magnitude $\min(M,d \min_i a_i)$. Apply Lemma \ref{nofalsedelscond} with $S_T = S_0 + f$, $S_\Delta = S_a$, $b_1 = \min(M,d \min_i a_i)$. It is applicable since conditions \ref{sone_sinf_bnd} and \ref{nofd_const} hold. Thus none of the constant coefficients will get falsely deleted and so $|\tDelta| \le S_a$. Also, clearly $|\tDelta_e| \le |\tDelta_{e,\dett}| \le f$. Thus $|\tT| \le S_0+f$.

For the induction step, assume that the result holds for $t_j+i-1$. Thus, at $t=t_j+i$, $|\Delta_{e,t}| = |\tDelta_{e,t-1}| \le if$ and $|\Delta_t| = |\tDelta_{t-1}| \le S_a$. Using condition \ref{algo_thresh}, after the detection step, $|\tDelta_{e,\dett}| \le (i+1) f$. Thus, $|\tT_\dett|  \le S_0 + (i+1)f$. Also, $|\tDelta_\dett| \le S_a$ and so $\|x_{\tDelta_\dett}\|_\infty \le \min(M, (i+1) (a_\add)_{(1)})$. Applying Lemma \ref{nofalsedelscond} with $S_T = S_0 + (i+1)f$, $S_\Delta = S_a$, $b_1 = \min(M,d \min_i a_i)$ (applicable since conditions \ref{sone_sinf_bnd}, \ref{nofd_const} hold), none of the constant coefficients will get falsely deleted. Thus, $|\tDelta| \le S_a$.
Also, clearly $|\tDelta_e| \le |\tDelta_{e,\dett}| \le (i+1)f$. Thus $|\tT| \le S_0+(i+1)f$.

\begin{claim}
At $t=t_j+d_0+i-1$, for all $i=1, \dots S_a$, $|\tDelta| \le S_a - i$, $|\tDelta_e| \le (d_0+i) f$, $|\tT| \le S_0+ (d_0+i) f$, and the first $i$ largest increasing coefficients are definitely detected.
\end{claim}

{\em Proof: } We prove this by induction. Consider the base case $t=t_j+d_0$. Using the previous claim, $|\Delta| \le S_a$, $|\Delta_e| \le d_0 f$, $|T| \le S_0 + d_0 f$. At this time, either the largest element of $\add$, which has magnitude $\min(M, (d_0+1) (a_\add)_{(1)})$, has already been detected so that the number of undetected elements  already satisfies $|\Delta| \le S_a-1$ or it has not been detected. If it has been detected, then $|\tDelta_\dett| \le |\Delta| \le S_a-1$. If it has not been detected, then $(x_\Delta)_{(1)} = \min(M, (d_0+1) (a_\add)_{(1)})$. Apply Lemma \ref{largest} with $S_\Delta=S_a$, $S_T =S_0 + d_0 f$. It is applicable since conditions \ref{sone_sinf_bnd}, \ref{thetabnd} hold and condition \ref{add_incr} holds for $i=1$. Thus the largest element will definitely get detected. Thus, in all cases, $|\tDelta_\dett| \le S_a-1$ and so $\|x_{\tDelta_\dett}\|_\infty \le \min(M, (d_0+1) (a_\add)_{(2)})$. Using condition \ref{algo_thresh}, $|\tDelta_{e,\dett}| \le (d_0+1)f$ and so $|\tT_\dett| \le S_0 + (d_0+1)f$.

Applying Lemma \ref{nofalsedelscond} with $S_T = S_0 + (d_0+1)f$, $S_\Delta = S_a-1$, $b_1 = \min(M,(d_0+1) (a_\add)_{(1)}$  (applicable since condition \ref{sone_sinf_bnd} holds and \ref{nofd_incr} holds for $i=1$), the largest increasing coefficient will not get falsely deleted. Further, applying Lemma \ref{nofalsedelscond} with $b_1 = \min(M,d \min_i a_i)$ (applicable since conditions \ref{sone_sinf_bnd} and \ref{nofd_const} hold), none of the constant coefficients will get falsely deleted.  Thus, $|\tDelta| \le S_a-1$. Also $|\tDelta_e| \le |\tDelta_{e,\dett}| \le (d_0+1)f$ and so $|\tT| \le S_0+(d_0+1)f$.

For the induction step, assume that the result holds for $t_j+d_0+i-2$. Thus, at $t=t_j+d_0+i-1$, $|\Delta| \le S_a - i+1$, $|\Delta_e| \le (d_0+i-1) f$, $|T| \le S_0+ (d_0+i-1) f$ and the first $i-1$ largest elements have already definitely been detected. Either the $i^{th}$ largest element has also been already detected, in which case $|\Delta| \le S_a - i$ or it has not been detected. If it has, then $|\tDelta_\dett| \le |\Delta| \le S_a-i$. If it has not been detected, then $(x_\Delta)_{(1)} = \min(M,(d_0+i) (a_\add)_{(i)})$. As before, use conditions \ref{sone_sinf_bnd}, \ref{thetabnd} and \ref{add_incr} and apply Lemma \ref{largest} to claim that the $i^{th}$ largest element will definitely get detected. Thus, in all cases, $|\tDelta_\dett| \le S_a-i$ and so $\|x_{\tDelta_\dett}\|_\infty \le \min(M, (d_0+i) (a_\add)_{(i+1)})$. Using condition \ref{algo_thresh}, $|\tDelta_{e,\dett}| \le (d_0+i)f$ and so $|\tT_\dett| \le S_0+(d_0+i)f$.
Also as before, apply Lemma \ref{nofalsedelscond} first with $b_1 = \min(M,(d_0+i) (a_\add)_{(i)})$ and then with $b_1 = \min(M,d \min_i a_i)$ (applicable since conditions \ref{sone_sinf_bnd}, \ref{nofd_incr},  \ref{nofd_const} hold) to claim that all constant coefficients and all the $i$ largest increasing coefficients will not get falsely deleted. Thus $|\tDelta| \le S_a-i$. Also,  $|\tDelta_e| \le (d_0+i)f$ and $|\tT| \le S_0+(d_0+i)f$.

\begin{claim}
At $t=t_j+d_0+S_a-1$, $|\tDelta_e|=0$.
\end{claim}

{\em Proof: } In the previous proof we have shown that at $t=t_j+d_0+S_a-1$, i.e. for $i=S_a$, $|\tDelta_\dett| = 0$ and $|\tT_\dett| \le S_0 + (d_0+S_a)f$. Apply Lemma \ref{truedelscond} with $S_\Delta=0$, $S_T =S_0 + (d_0+S_a)f$
(applicable since conditions \ref{sone_sinf_bnd}, \ref{algo_thresh} hold). Thus all false detects will get deleted, i.e.  $|\tDelta_e|=0$.

\begin{claim}
At all $t \in [t_j+d_0+S_a-1, t_{j+1}-r-1]$, $|\tDelta|=0$,  $|\tDelta_e|=0$. Thus  $\tT = N_t$ and  $|\tT| = |N_t|=S_0$.
\end{claim}

{\em Proof: } Using the previous two claims, the result holds for $t=t_j+d_0+S_a-1$ (base case). For the induction step, assume that it holds for $t_j+d_0+S_a+i-1$. Thus, at $t=t_j+d_0+S_a+i$, $|\Delta|=0$, $|\Delta_e|=0$ and $|T|=S_0$. Since $|\tDelta_{\dett}| \le |\Delta|$, $|\tDelta_{\dett}|=0$ and thus 
 $\|x_{\tDelta_\dett}\|_\infty = 0$. Using condition \ref{algo_thresh}, $|\tDelta_{e,\dett}| \le 0+f$ and thus $|\tT_\dett| \le S_0+f$. Use conditions \ref{sone_sinf_bnd}, \ref{nofd_incr} (for $i=S_a$) and \ref{nofd_const} to first apply Lemma \ref{nofalsedelscond} with $S_T=S_0+f$, $S_\Delta=0$, $b_1 =  \min(M, (d_0+S_a+i+1) (a_\add)_{(S_a)}$ (smallest increasing coefficient) and then with $b_1= \min(M,d \min_i a_i)$ (smallest constant coefficient) to show that there are no false deletions of either constant or increasing coefficients. Thus $|\tDelta|=0$. Use conditions \ref{sone_sinf_bnd} and \ref{algo_thresh} and apply Lemma \ref{truedelscond} with $S_\Delta=0$, to show that $|\tDelta_e|=0$.

\begin{claim}
At $t \in [t_{j+1}-r,t_{j+1}-1]$, $|\tDelta|=0$,  $|\tDelta_e|=0$. Thus  $\tT = N_t$ and  $|\tT| = |N_t|=S_0$.
\end{claim}

{\em Proof: } The proof again follows by induction and arguments similar to those of the previous claim. The only difference is the following. At any $t=t_{j+1}-r+i-1$, one applies Lemma \ref{nofalsedelscond} three times: the first two times for increasing and constant coefficients (as before) and then a third time with $S_T=S_0+f$, $S_\Delta=0$, $b_1 = ((i-1)/r)\min(M,d (a_\rem)_{(S_a)})$
(for the current smallest decreasing coefficient). This last one is applicable since conditions \ref{sone_sinf_bnd} and \ref{nofd_decr} hold.

\begin{claim}
At $t = t_{j+1}-1$,  $|\tDelta|=0$,  $|\tDelta_e|=0$. Thus  $\tT = N_t$ and  $|\tT| = |N_t|=S_0-S_a$.
\end{claim}
The only difference at this time is that the decreasing coefficients get removed. As a result, $|N_t|=S_0-S_a$, $|\Delta_e|=S_a$ and $|\Delta_{e,\dett}|=S_a+f$. But $|\Delta_\dett| = |\Delta|=0$. As before, using conditions \ref{sone_sinf_bnd} and \ref{algo_thresh} and applying Lemma \ref{truedelscond} with $S_\Delta=0$, all extras will still get removed and so still $|\tDelta_e|=0$. Everything else is the same as before.


\subsection{LS-CS Stability: Proof of Corollary \ref{stabres_cor}}
\label{corproof}

We have shown that $|\tT| \le S_0 + f(d_0+S_a)$ and $|\tDelta| \le S_a$. We can bound $\|x_\tDelta\|$ as follows.
In the first sub-interval, $|\tDelta| \le S_a$ and the maximum value of any element of $\tDelta_t$ at any $t$ in this interval is $\min(M,d_0 (a_{\add(j)})_{(1)})$ so that $\|x_\tDelta\|^2 \le S_a \min(M,d_0 (a_{\add(j)})_{(1)})^2$.
In the second sub-interval, at $t=t_j+d_0+i-1$, $|\tDelta| \le S_a-i$ and $\|x_\tDelta\|_\infty \le \min(M, (d_0+i) (a_{\add(j)})_{(i+1)})$. In the last two sub-intervals, $|\tDelta|=0$. Thus,
\bea
\|x_\tDelta\|^2 \sle \max_{j} \max_{i=0, \dots S_a} (S_a-i) \min(M, (d_0+i) (a_{\add(j)})_{(i+1)})^2 \nn \\
 \sle S_a  \min(M, (d_0+S_a) \max_i a_i)^2
\label{xdelta_bnd}
\eea
This gives the LS-CS error bound.
In a similar fashion, we can argue that $\|x_\Delta\|^2 \le S_a  \min(M, (d_0+S_a) \max_i a_i)^2$. Using this in Corollary \ref{cor1} gives the CS-residual error bound.

\bibliographystyle{IEEEbib}
\bibliography{tipnewpfmt_kfcsfullpap,tipnewpfmt} 








\end{document}